\documentclass[lettersize,journal]{IEEEtran}
\usepackage{amsmath,amsfonts}
\usepackage{algorithmic}
\usepackage{algorithm}
\usepackage{array}
\usepackage[caption=false,font=normalsize,labelfont=sf,textfont=sf]{subfig}
\usepackage{textcomp}
\usepackage{stfloats}
\usepackage{url}
\usepackage{verbatim}
\usepackage{graphicx}
\usepackage{booktabs} 
\usepackage{nicefrac} 
\usepackage{microtype}
\usepackage{xcolor} 
\usepackage{float}
\usepackage{cite}
\usepackage{multirow}
\usepackage{hyperref}
\begin{document}

\title{SoloSpeech: Enhancing Intelligibility and Quality in Target Speech Extraction through a Cascaded Generative Pipeline}

\author{Helin Wang, Jiarui Hai, Dongchao Yang, Chen Chen, Kai Li, Junyi Peng, Thomas Thebaud,~\IEEEmembership{Member,~IEEE}, Laureano Moro-Velázquez,~\IEEEmembership{Member,~IEEE}, Jesús Villalba,~\IEEEmembership{Member,~IEEE}, and Najim Dehak,~\IEEEmembership{Senior~Member,~IEEE}
\thanks{
Helin Wang, Jiarui Hai, Thomas Thebaud, Laureano Moro-Velázquez, Jesús Villalba and Najim Dehak are with the Johns
Hopkins University (email: $\{$hwang258, jhai2, tthebau1, laureano, jvillal7, ndehak3$\}$@jhu.edu).}
\thanks{
Dongchao Yang is with the Chinese University of Hong Kong (email: dcyang@se.cuhk.edu.hk). Chen Chen is with the NVIDIA (email: cchen1@nvidia.com).
Kai Li is with the Tsinghua University (email: tsinghua.kaili@gmail.com).
Junyi Peng is with the Brno University of Technology (email: pengjy@fit.vut.cz).
}}



\maketitle

\begin{abstract}
  Target Speech Extraction (TSE) aims to isolate a target speaker's voice from a mixture of multiple speakers by leveraging speaker-specific cues, typically provided as auxiliary audio (\textit{a.k.a.} cue audio). Although recent advancements in TSE have primarily employed discriminative models that offer high perceptual quality, these models often introduce unwanted artifacts, reduce naturalness, and are sensitive to discrepancies between training and testing environments. On the other hand, generative models for TSE lag in perceptual quality and intelligibility. To address these challenges, we present \textbf{SoloSpeech}, a novel cascaded generative pipeline that integrates compression, extraction, reconstruction, and correction processes. SoloSpeech features a speaker-embedding-free target extractor that utilizes conditional information from the cue audio's latent space, aligning it with the mixture audio's latent space to prevent mismatches. Evaluated on the widely-used Libri2Mix dataset, SoloSpeech achieves the new state-of-the-art intelligibility and quality in TSE while demonstrating exceptional \textbf{generalization on out-of-domain data} and \textbf{real-world scenarios}. 
\end{abstract}


\section{Introduction}
\label{intro}
Humans possess the remarkable ability to focus on a specific speaker amidst a cacophony of sounds, including background noise, reverberation, and interfering voices. This capability is famously referred to as the cocktail-party effect \cite{bronkhorst2015cocktail}. For many years, researchers have endeavored to emulate this sophisticated auditory skill using algorithms. 
Target Speech Extraction (TSE) emerges as a promising field, aiming to isolate a desired speaker's voice from a complex mixture of multiple speakers, alongside any accompanying noise or reverberation \cite{DBLP:journals/spm/ZmolikovaDOKCY23}. TSE leverages cues, which serve as identifiers for the target speaker, to accurately extract their speech from a mixture. 
Various types of cues have been investigated to address the TSE problem. These include spatial cues \cite{DBLP:conf/interspeech/GuCZZXYSZ019,DBLP:conf/icassp/SubramanianWYZX20}, visual cues \cite{DBLP:journals/tog/EphratMLDWHFR18}, concept cues \cite{DBLP:conf/mm/OhishiDOATNKHK22}, and audio cues \cite{DBLP:conf/interspeech/WangMWSWHSWJL19,DBLP:journals/jstsp/ZmolikovaDKONBC19,DBLP:conf/interspeech/WangZLLWG0Q024}. In this wor, we specifically focus on monaural TSE methods that utilize the most prevalent audio-based cues.

A significant challenge in TSE lies in distinguishing between target and non-target speakers, as their speech signals often share similar characteristics, complicating the extraction process \cite{DBLP:conf/interspeech/ZhaoYGZZ22}. To overcome this, many researchers have concentrated on discriminative model-based TSE, which directly maps the mixture signal to the target signal by optimizing signal-level metrics \cite{han2022dpccn, DBLP:conf/interspeech/WangYWYZ22, DBLP:journals/corr/abs-2409-15799}. Both time-domain \cite{DBLP:journals/taslp/XuRCL20,DBLP:conf/interspeech/GeXWCD020,DBLP:conf/icassp/LiuDWZ23} and time-frequency (T-F) domain methodologies \cite{DBLP:journals/inffus/HaoLZ24,DBLP:journals/corr/abs-2409-02615} have been investigated. While these discriminative approaches are highly effective, they can sometimes introduce artifacts and are often sensitive to discrepancies between training and testing conditions \cite{kamo23_interspeech,wang2024noise}.
In contrast, generative approaches model the joint distribution of the mixture speech, the target speech, and the cue speech. 
Various deep generative models have been explored for tasks such as speech enhancement, dereverberation, and separation \cite{DBLP:journals/taslp/RichterWLLG23,scheibler2023diffusion,DBLP:conf/iclr/Liu0VSTH24,DBLP:journals/corr/abs-2502-03128}. These models have demonstrated the ability to exhibit greater robustness in unseen scenarios than discriminative models \cite{DBLP:journals/taslp/RichterWLLG23,wang2024noise}. They are capable of achieving high clarity in overlapping regions where multiple speakers converse simultaneously \cite{DBLP:journals/corr/abs-2409-08425}. However, current generative approaches still result in separated speech with reduced audio quality and intelligibility, leading mainstream TSE solutions to continue relying on discriminative methods \cite{DBLP:journals/corr/abs-2301-10752}.

To \textbf{consistently enhance the quality and intelligibility of generative models} for TSE, we introduce SoloSpeech\footnote{Audio samples, source code and pretrained models are available at \href{https://wanghelin1997.github.io/SoloSpeech-Demo/}{https://wanghelin1997.github.io/SoloSpeech-Demo/}},
a novel cascaded generative pipeline that comprises three key components: a generative audio compressor, a generative target extractor, and a generative corrector. To the best of our knowledge, we are the first to introduce such a pipeline for target speech extraction. 
The audio compressor utilizes a T-F domain variational autoencoder (VAE) to transform audio waveforms into latent representations and vice versa. The target extractor employs a latent diffusion model to derive the latent representation of the target signal. Notably, we propose a speaker-embedding-free model for the target extractor, where the mixture audio and the cue audio (\textit{i.e.} condition) are fused using a cross-attention mechanism \cite{vaswani2017attention} in the same latent space, preventing potential mismatches.
Furthermore, inspired by recent research on error correction with generative models \cite{wang2024noise,DBLP:journals/corr/abs-2409-09785, mu2025sepalm}, we propose a T-F domain diffusion model as the corrector. This model addresses errors introduced by the target extractor and refines the audio quality, ensuring high quality and  intelligibility in the extracted speech.
We evaluated the proposed pipeline on the Libri2Mix noisy benchmark dataset \cite{cosentino2020librimix} and demonstrated its significant superiority over existing methods in the TSE task. SoloSpeech achieves new state-of-the-art results in perceptual quality, naturalness, and intelligibility, underscoring its effectiveness and robustness in noisy acoustic scenarios. Moreover, experiments on three out-of-domain datasets and two real-word datasets showed that SoloSpeech exhibits \textbf{strong generalization capabilities on unseen conditions}.

\section{Related Work}
\label{relatedw}
\subsection{Generative models in TSE}
\label{relatedw1}
Recent works \cite{DBLP:conf/interspeech/KamoDN23,hai2024dpm} have introduced TSE systems utilizing diffusion probabilistic models, with similar methods being explored for SS \cite{DBLP:conf/icassp/ScheiblerJCBCC23}. SpeechFlow \cite{DBLP:conf/iclr/Liu0VSTH24} proposed a generative model pretrained on large-scale speech data using flow matching and masked conditions, and Ku, et al.\ (2024) \cite{DBLP:journals/corr/abs-2409-16117} adapted this method for TSE.
Additionally, Yu, et al.\ (2024) \cite{yu2024generation} introduced a discrete token-based TSE approach by combining speech discretization with vocoder techniques and Ma, et al.\ (2025) \cite{DBLP:journals/corr/abs-2501-14477} adopted the Whisper audio encoder to improve the intelligibility of generative TSE.
These models are capable of generating audio samples with high naturalness and demonstrate good generalization to out-of-domain data. However, they often yield inferior results on reference-based metrics \cite{DBLP:journals/corr/abs-2305-05857}.
To further enhance perceptual quality, researchers have begun integrating generative models with discriminative models \cite{DBLP:journals/taslp/LemercierRWG23,wang2024noise,DBLP:journals/corr/abs-2301-10752,zhang2024ddtse}. 
In contrast, our proposed SoloSpeech employs a \textbf{fully generative approach}, achieving superior \textbf{perceptual quality, naturalness, and intelligibility}.

\subsection{Conditional information in TSE}
\label{relatedw2}
Conditional information plays a pivotal role in TSE by isolating the target speaker and attending to the mixture audio through various fusion layers, such as concatenation \cite{DBLP:journals/tog/EphratMLDWHFR18}, multiplication \cite{DBLP:journals/jstsp/ZmolikovaDKONBC19}, and factorized layers \cite{DBLP:conf/interspeech/ZmolikovaDKHON17,DBLP:journals/jstsp/ZmolikovaDKONBC19}. Existing methods for extracting conditional information can be categorized into speaker embedding-based approaches and speaker embedding-free approaches. Speaker embedding-based approaches include SpEx+ \cite{DBLP:conf/interspeech/GeXWCD020}, which trained speaker embeddings from scratch using an auxiliary speaker identification task, X-SepFormer \cite{DBLP:conf/icassp/LiuDWZ23}, built upon x-vectors \cite{DBLP:conf/icassp/SnyderGSPK18}, X-TF-GridNet \cite{DBLP:journals/inffus/HaoLZ24}, which utilized adaptive speaker embedding fusion. \cite{10627502,10448315} used pre-trained self-supervised learning (SSL) models to derive speaker embeddings. Conversely, speaker embedding-free approaches such as SMMA-Net \cite{DBLP:conf/icassp/HuXGHH24} and CIENet \cite{DBLP:conf/icassp/YangBZC24} employed attention mechanisms to interact with the representations of the reference and mixed signals, VE-VE \cite{DBLP:conf/icassp/YangLTYM23} used an RNN-based voice extractor to capture speaker characteristics, and SEF-Net \cite{DBLP:conf/interspeech/ZengSW023} along with USEF-TSE \cite{DBLP:journals/corr/abs-2409-02615} utilized cross-attention to implicitly leverage speaker information. 
Our proposed method adopts a speaker embedding-free approach, eliminating the need for a pre-trained model to extract fixed-length speaker embeddings or align the speaker embedding space with the target latent space. Compared to speaker embedding-based methods, our approach \textbf{preserves local dynamics and temporal structures}, which are crucial for guiding more effective speaker extraction.

\section{Diffusion probabilistic model}
\label{ap1}
Diffusion Probabilistic Models (DPMs) include a forward and a backward process. In the forward process, Gaussian noise is progressively added to the data following a pre-defined variance schedule \(\beta_{1}, \ldots, \beta_{T}\). 
\begin{equation}
q\left(x_{1: T} \mid x_{0}\right):=\prod_{t=1}^{T} q\left(x_{t} \mid x_{t-1}\right) \\
\end{equation}
\begin{equation}
q\left(x_{t} \mid x_{t-1}\right):=\mathcal{N}\left(x_{t} ; \sqrt{1-\beta_{t}} x_{t-1}, \beta_{t} \mathbf{I}\right)
\end{equation}

The forward process allows sampling \(x_{t}\) at an arbitrary timestep \(t\) in a closed form:
\begin{equation}
q\left(x_{t} \mid x_{0}\right):=\mathcal{N}\left(x_{t} ; \sqrt{\bar{\alpha}_{t}} x_{0},\left(1-\bar{\alpha}_{t}\right) \mathbf{I}\right)
\end{equation}
Equivalently:
\begin{equation}
\label{eq10}
x_{t}:=\sqrt{\bar{\alpha}_{t}} x_{0}+\sqrt{1-\bar{\alpha}_{t}} \xi, \quad \text { where } \xi \sim \mathcal{N}(\mathbf{0}, \mathbf{I})
\end{equation}
where \(\alpha_{t}:=1-\beta_{t}\) and \(\bar{\alpha}_{t}:=\) \(\prod_{s=1}^{t} \alpha_{s}\). 

DPMs learn the reverse process to recover information incrementally. In this manner, DPMs can generate new data from random Gaussian noise. When \(\beta_{t}\) is small, the reverse step is also Gaussian:

\begin{equation}
p_{\theta}\left(x_{0: T}\right):=p\left(x_{T}\right) \prod_{t=1}^{T} p_{\theta}\left(x_{t-1} \mid x_{t}\right)
\end{equation}
\begin{equation}
p_{\theta}\left(x_{t-1} \mid x_{t}\right):=\mathcal{N}\left(x_{t-1} ; \tilde{\mu}_{t}, \tilde{\beta}_{t} \mathbf{I}\right)
\end{equation}
where variance \(\tilde{\beta}_{t}\) can be calculated using the forward process posteriors:
$\tilde{\beta}_{t}:=\frac{1-\bar{\alpha}_{t-1}}{1-\bar{\alpha}_{t}} \beta_{t}$

Neural networks are typically used to predict noise \(\xi\), since:
\begin{equation}
\label{eq13}
\tilde{\mu}_{t}:=\frac{1}{\sqrt{\alpha_{t}}}\left(x_{t}-\frac{\beta_{t}}{\sqrt{1-\bar{\alpha}_{t}}} \xi\right)
\end{equation}

Following \cite{DBLP:conf/wacv/LinLLY24},
we introduce a modified noise schedule with zero terminal SNR, where the SNR can be calculated as:

\begin{equation}
\operatorname{SNR}(t):=\frac{\bar{\alpha}_{t}}{1-\bar{\alpha}_{t}}
\end{equation}


We adjust the noise schedule to enforce zero terminal $\operatorname{SNR}(T)$  by keeping \(\sqrt{\bar{\alpha}_{1}}\) unchanged, changing \(\sqrt{\bar{\alpha}_{T}}\) to zero, and linearly rescaling \(\sqrt{\bar{\alpha}_{t}}\) for intermediate \(t \in[2, \ldots, T-1]\) respectively.

When the SNR reaches zero at the terminal step, predicting noise 
\(\xi\) becomes meaningless because the input and output are identical. Consequently, the neural network is modified to predict velocity $v_t$ instead, as present in \cite{DBLP:conf/iclr/SalimansH22}:

\begin{equation}
v_{t}:=\sqrt{\bar{\alpha}_{t}} \xi-\sqrt{1-\bar{\alpha}_{t}} x_{0}
\end{equation}
\begin{equation}
\xi=\sqrt{\bar{\alpha}_t} v_t+\sqrt{1-\bar{\alpha}_t} x_t
\end{equation}

According to (\ref{eq10}) and (\ref{eq13}), the backward process is performed by the following functions:
\begin{equation}
x_{0}:=\sqrt{\bar{\alpha}_{t}} x_{t}-\sqrt{1-\bar{\alpha}_{t}} v_{t}
\end{equation}
\begin{equation}
\tilde{\mu}_{t}:=\frac{\sqrt{\bar{\alpha}_{t-1}} \beta_{t}}{1-\bar{\alpha}_{t}} x_{0}+\frac{\sqrt{\alpha_{t}}\left(1-\bar{\alpha}_{t-1}\right)}{1-\bar{\alpha}_{t}} x_{t}
\end{equation}

At the terminal timestep, the neural network trained to predict velocity estimates the mean of the data distribution under the given conditions. The diffusion sampler consistently initiates from the final timestep during inference.

\begin{figure}[t]
\begin{center}
\centerline{\includegraphics[width=\columnwidth]{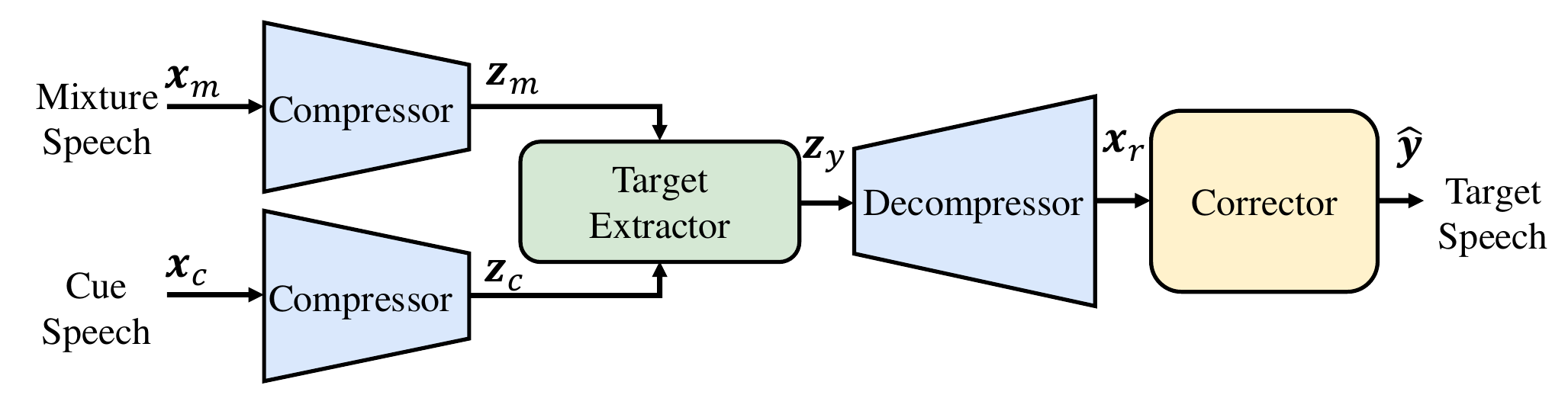}}
\caption{Overall pipeline of SoloSpeech.}
\label{pic0}
\end{center}
\end{figure}

\section{SoloSpeech}
\label{solospeech}
In this section, we present a cascaded generative pipeline that operates through a sequence of compression, extraction, reconstruction, and correction stages, which is shown in Figure~\ref{pic0}. 

\subsection{Overall pipeline}
 Let $\boldsymbol{x}_m \in \mathbb{R}^{1 \times T_m}$ and $\boldsymbol{x}_c \in \mathbb{R}^{1 \times T_c}$ denote the mixture speech and the conditional speech (\textit{a.k.a.} cue speech), respectively, where $T_m$ and $T_c$ represent their corresponding audio lengths. The task of TSE is to estimate the target speech signal $\boldsymbol{y} \in \mathbb{R}^{1 \times T_m}$.

The audio compressor transforms the speech signals into compact latent representations: $\boldsymbol{z}_m \in \mathbb{R}^{D \times L_m}$ for the mixture speech and $\boldsymbol{z}_c \in \mathbb{R}^{D \times L_c}$ for the conditional speech (see appendix~\ref{solospeech1} for details). Here, $D$ denotes the dimensionality of the latent features for each frame, while $L_m$ and $L_c$ represent the number of frames in the mixture and conditional latents, respectively.
The target extractor (detailed in appendix~\ref{solospeech2}) then predicts the latent features of the target speech, $\boldsymbol{z}_y \in \mathbb{R}^{D \times L_m}$, from $\boldsymbol{z}_m$ conditioned on $\boldsymbol{z}_c$. The estimated target speech signal, $\boldsymbol{x}_r \in \mathbb{R}^{1 \times T_m}$, is reconstructed using the decoder of the audio compressor.
Finally, we introduce a corrector to refine the target speech, resulting in the refined signal $\boldsymbol{\hat{y}} \in \mathbb{R}^{1 \times T_m}$. See appendix~\ref{solospeech3} for details.

\subsection{Audio compressor}
\label{solospeech1}
\begin{figure}[t]
\begin{center}
\centerline{\includegraphics[width=0.8\columnwidth]{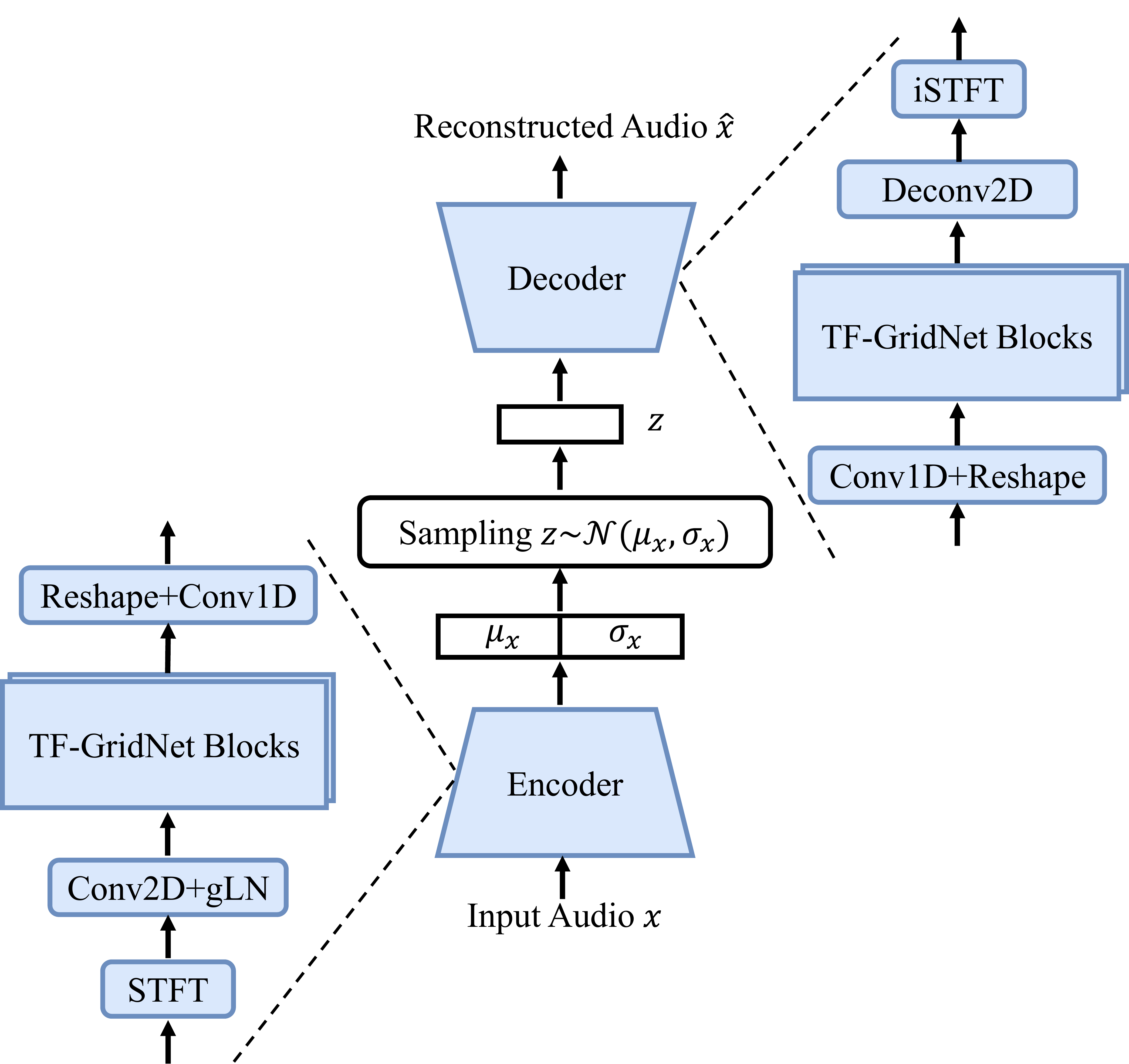}}
\caption{The audio compressor architecture.}
\label{pic1}
\end{center}
\end{figure}

The audio compressor operates on raw waveforms, compressing them into manageable sequence lengths. Current advanced audio compressors utilize time-domain VAEs with several convolutional blocks \cite{DBLP:conf/nips/KumarSLKK23,DBLP:journals/corr/abs-2407-14358,hai2024ezaudio}.
Inspired by the success of T-F domain modeling for mixture speech signals \cite{DBLP:journals/taslp/WangCCLKW23,DBLP:journals/corr/abs-2404-02063}, we propose a \textbf{novel T-F domain VAE} as the audio compressor. As illustrated in Figure~\ref{pic1}, the encoder first applies the Short-Time Fourier Transform (STFT) \cite{DBLP:journals/taslp/WilliamsonWW16} to transform the input audio signal $\boldsymbol{x} \in \mathbb{R}^{1 \times N}$ into a complex spectrum $\boldsymbol{S} \in \mathbb{R}^{2 \times F \times \frac{N}{h}}$, where $N$ denotes the audio length, $h$ the hop size of STFT, and $F$ the number of frequency bins. The real and imaginary components are concatenated along the first dimension.

Our VAE adopts TF-GridNet \cite{DBLP:journals/taslp/WangCCLKW23} as its backbone network, which comprises several TF-GridNet blocks. The encoder outputs a latent representation of shape $\mathbb{R}^{1 \times 2D \times \frac{N}{h}}$, which is then split into two parts: the mean $\mu_x \in \mathbb{R}^{1 \times D \times \frac{N}{h}}$ and the variance $\sigma_x \in \mathbb{R}^{1 \times D \times \frac{N}{h}}$. The latent representation is obtained by sampling from this distribution, denoted as $z \sim \mathcal{N}(\mu_x, \sigma_x)$ where $z \in \mathbb{R}^{1 \times D \times \frac{N}{h}}$.

The decoder mirrors the encoder structure, reconstructing the audio waveform using the inverse STFT (iSTFT). The VAE model is trained in a generative and adversarial manner \cite{DBLP:journals/corr/abs-2407-14358}. The loss functions include: (i) a perceptually weighted multi-resolution STFT reconstruction loss \cite{steinmetz2020auraloss}; (ii) an adversarial loss term with feature matching, utilizing five convolutional discriminators as in Encodec \cite{DBLP:journals/tmlr/DefossezCSA23}; and (iii) a KL divergence loss term. See appendix~\ref{ap41} for details.

\subsection{Target extractor}
\label{solospeech2}
\begin{figure*}[t]
  \centering
  \subfloat[]{\includegraphics[width=0.45\linewidth]{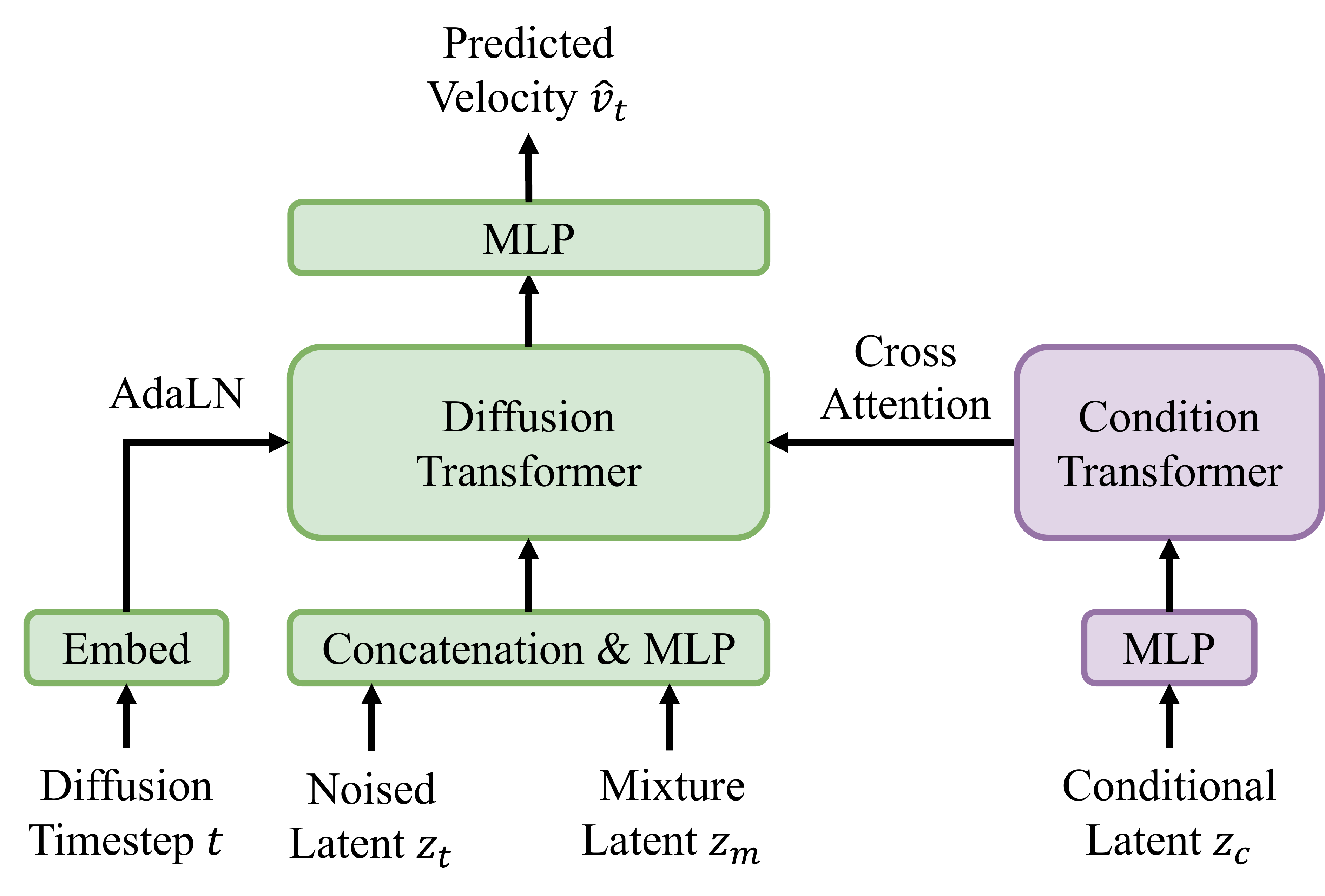}\label{pic2}}
\hfill
\subfloat[]{\includegraphics[width=0.25\linewidth]{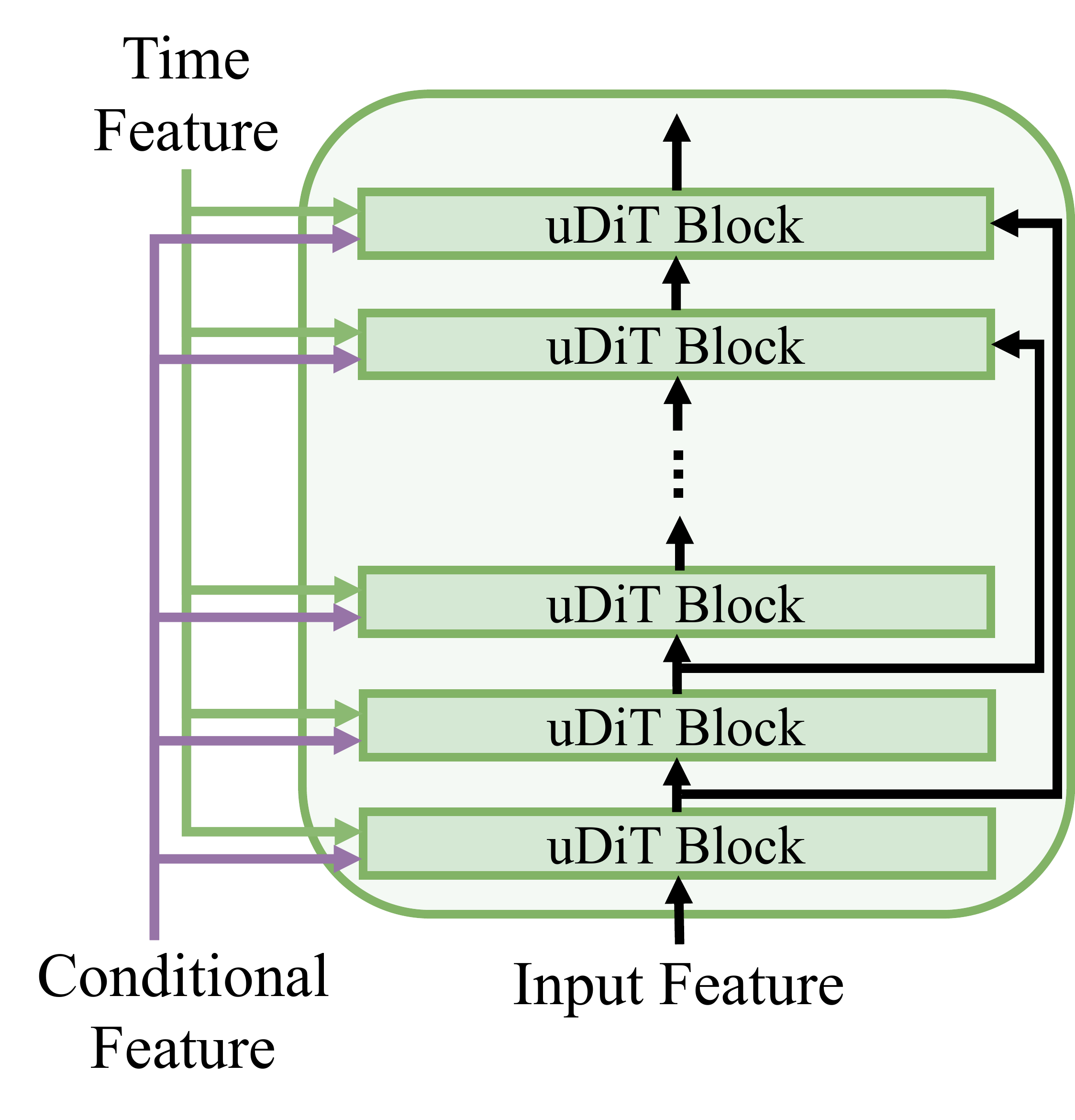}\label{fig:sub1}}
\hfill
\subfloat[]{\includegraphics[width=0.25\linewidth]{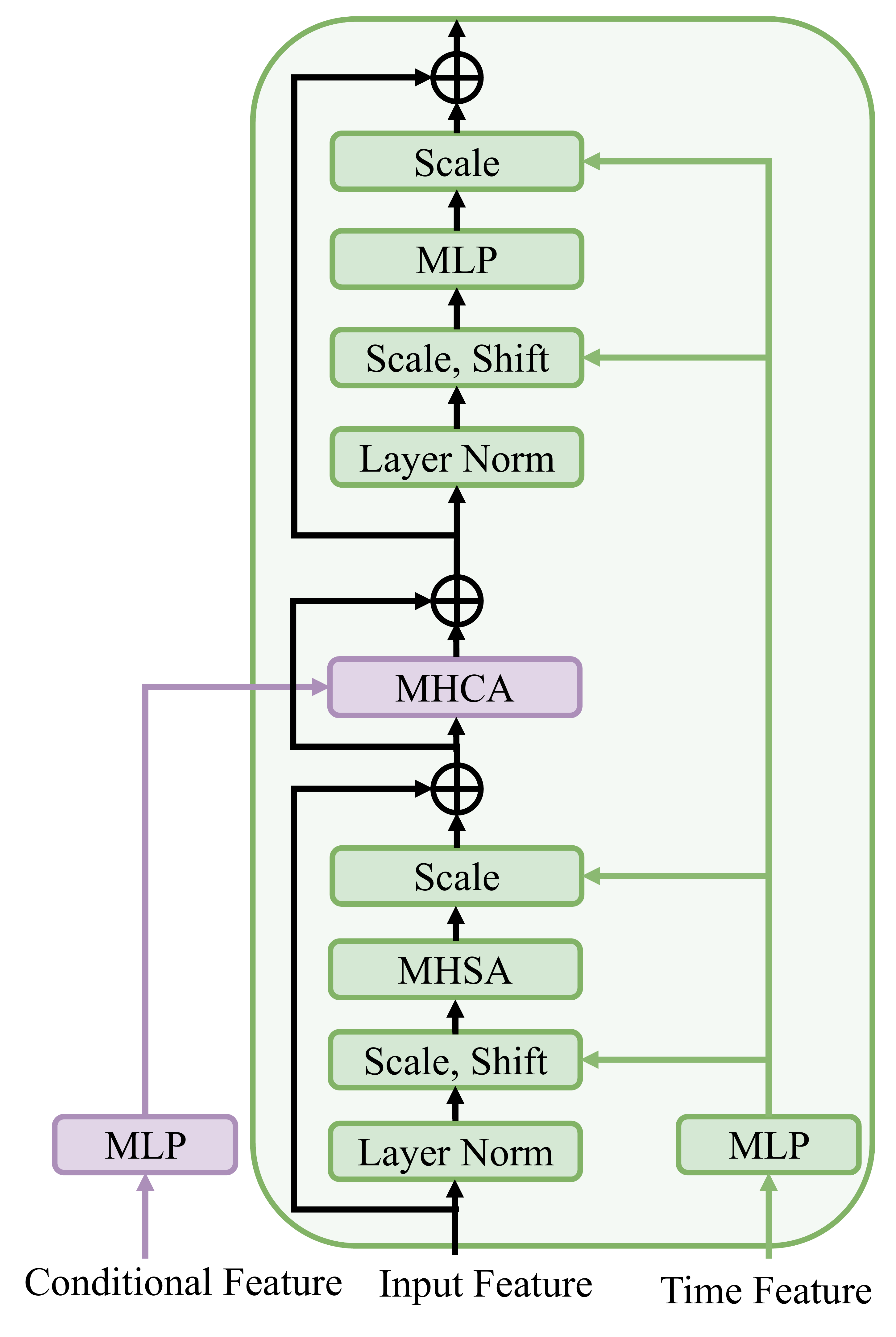}\label{fig:sub2}}
\caption{Architectures of the target extractor (a), Diffusion Transformer backbone (b) and uDiT block (c).}
\label{fig:main}
\end{figure*}

The target extractor operates on the latent representations, aiming to extract the target latents $\boldsymbol{z}_y$ from the mixture latents $\boldsymbol{z}_m$, guided by the conditional latents $\boldsymbol{z}_c$. As shown in Figure~\ref{pic2}, we use a modified diffusion scheduler and velocity prediction \cite{DBLP:conf/wacv/LinLLY24}, which have been shown to improve the purity and overall performance of sound extraction \cite{hai2024dpm}. Details of the diffusion process could be found in appendix~\ref{ap1}.
The main network for the target extractor employs a diffusion transformer with long skip connections (uDiT) \cite{DBLP:journals/corr/abs-2409-08425}, a modified version of DiT \cite{DBLP:conf/iccv/PeeblesX23}, as detailed in Figure~\ref{fig:sub1}. These skip connections bridge shallow and deep DiT blocks, creating shortcuts for low-level features and streamlining the training of the entire velocity-prediction network. The input feature is a concatenation of the current noisy latent $\boldsymbol{z}_t$ and the mixture latent $\boldsymbol{z}_m$ in the latent dimension ($D$). The time feature and the conditional feature are fused into each block.

Unlike previous works that introduce speaker embeddings as conditional information \cite{hai2024dpm,DBLP:journals/corr/abs-2409-08425}, our method jointly trains a condition transformer to extract the conditional features from the conditional latents $\mathbf{z}_c$, which are then attended to by the diffusion transformer using a cross-attention mechanism. This approach offers the following advantages: (i) it eliminates the need for \textbf{extra data or labels} to train a speaker embedding network; (ii) it preserves the \textbf{sequential information} of the conditional features, such as local dynamics and temporal structures; and (iii) it prevents \textbf{potential mismatching} between VAE latents and other types of features representing the target speech. 

The condition transformer employs a ViT-like backbone \cite{DBLP:conf/iclr/DosovitskiyB0WZ21}, consisting of several ViT blocks. Each ViT block contains two LayerNorm layers, a Multi-Head Self-Attention (MHSA) layer, and MLP layers with SiLU activations \cite{DBLP:journals/nn/ElfwingUD18}. 
The condition transformer generates a sequence of conditional features, which are then attended to by the uDiT blocks. As shown in Figure~\ref{fig:sub2}, each uDiT block includes an MHSA module, a Multi-Head Cross-Attention (MHCA) module, and an MLP module. The conditional features are fused by the MHCA module, and the embedded time features are integrated into the MHSA module and the MLP module using Adaptive LayerNorm (AdaLN) layers. SiLU serves as the activation function, and we incorporate rotary positional embeddings (RoPE) \cite{DBLP:journals/ijon/SuALPBL24} for the position encoding in both MHCA and MHSA.

Following the reverse process of diffusion models, the target extractor gradually reconstructs the target latents from random Gaussian noise, generating the estimated target latents $\boldsymbol{z}_y$. The speech signal is then obtained from the estimated target latents using the decoder of the audio compressor, resulting in the reference signal $\boldsymbol{x}_r \in \mathbb{R}^{1 \times T_m}$.

\subsection{Corrector}
\label{solospeech3}

Diffusion models have been shown to outperform discriminative models for non-additive corruption types or when evaluated under mismatched conditions \cite{scheibler2023diffusion}. However, they often yield worse results on reference-based metrics \cite{DBLP:journals/corr/abs-2305-05857}, particularly for latent diffusion models, as VAE reconstruction determines the upper bound of audio quality. 
In addition, diffusion models may produce vocalizing and breathing artifacts in adverse conditions \cite{DBLP:journals/taslp/LemercierRWG23}.

Inspired by the success of incorporating generative models to refine signal quality \cite{wang2024noise}, we propose a T-F domain diffusion model-based corrector to address these issues. The corrector is designed to: (i) \textbf{reduce artifacts} caused by the front-end models; (ii) improve the \textbf{signal quality} from the audio compressor's output by incorporating signal quality-based training objective functions; and (iii) \textbf{correct errors} introduced by the target extractor, such as speaker obfuscation and mispronunciation, thereby enhancing intelligibility.

\subsubsection{Fast-GeCo Corrector}

Inspired by Fast-GeCo \cite{wang2024noise}, we employ a single-step diffusion model that allows us to directly optimize signal quality using the Scale-Invariant Signal-to-Noise Ratio (SI-SNR) loss function \cite{DBLP:conf/interspeech/IsikRCWH16,DBLP:journals/taslp/LuoM19}. As illustrated in Figure~\ref{fig4:sub1}, Fast-GeCo initially trains a multi-step diffusion model, which takes a noised signal as input, conditioned on the reference signal, mixture signal, and timestep. The noised signal depends on both the timestep and the reference signal. 
Subsequently, Fast-GeCo accelerates the traditional multi-step diffusion process by distilling it into a single-step operation, thereby producing the refined signal $\hat{\boldsymbol{y}} \in \mathbb{R}^{1 \times T_m}$.
Given the ground-truth target signal $\boldsymbol{y}$, the signal-step corrector is optimized by:
\begin{equation}
\mathcal{L}:=-10 \log _{10} \frac{\left\|\frac{\left\langle\hat{\boldsymbol{y}}, \boldsymbol{y}\right\rangle \boldsymbol{y}}{\left\|\boldsymbol{y}\right\|^2}\right\|^2}{\left\|\hat{\boldsymbol{y}}-\frac{\left\langle\hat{\boldsymbol{y}}, \boldsymbol{y}\right\rangle \boldsymbol{y}}{\left\|\boldsymbol{y}\right\|^2}\right\|^2}
\end{equation}
where $\left\|\boldsymbol{y}\right\|^2 = \left\langle\boldsymbol{y}, \boldsymbol{y}\right\rangle$ denotes the signal power. Scale invariance is maintained
by normalizing $\hat{\boldsymbol{y}}$ and $\boldsymbol{y}$ to zero mean before computation.

\subsubsection{SoloSpeech Corrector}

From our initial experiments, we observe that such corrector can improve signal quality and reduce artifacts from the target extractor. However, it may struggle when processing reference signals with speaker obfuscation or mispronunciation. This can result in weak \textbf{speaker similarity} between the target and extracted speech, as well as poor performance in \textbf{Automatic Speech Recognition} (ASR) tasks. We argue that this issue arises because the reference signal already contains sufficient information for the corrector to learn the refinement task, leading to inadequate utilization of the original mixture signal. Consequently, the corrector behaves more like a Speech Enhancement (SE) tool rather than a TSE tool.

\begin{figure}[t]
  \centering
  \subfloat[]{\includegraphics[width=0.9\linewidth]{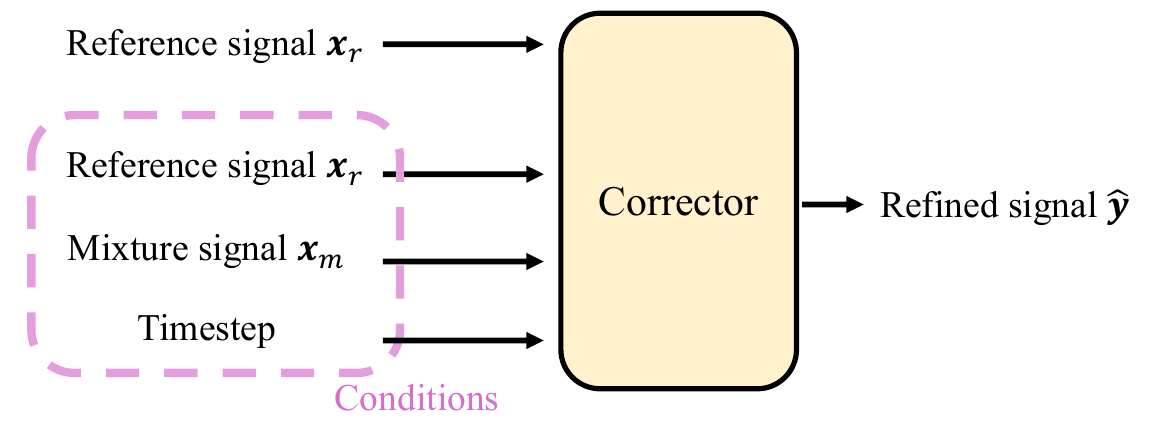}\label{fig4:sub1}}
  \hfill
  \subfloat[]{\includegraphics[width=0.9\linewidth]{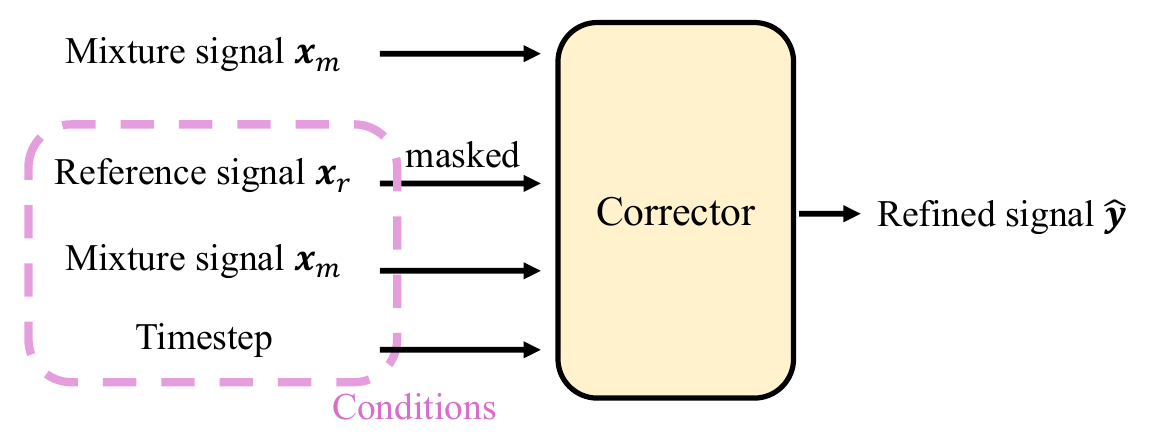}\label{fig4:sub2}}
  \caption{Diagrams of Fast-GeCo corrector (a) and SoloSpeech corrector (b).}
  \label{pic4}
\end{figure}

To address the above issues, in SoloSpeech, the refined signal is derived from the mixture signal rather than the reference signal, \textbf{enabling the corrector to function similarly to a TSE task}. As illustrated in Figure~\ref{fig4:sub2}, the corrector receives a noised mixture signal as input.
To encourage the corrector to leverage more information from the mixture signal and enhance the intelligibility of the refined signal, we propose a simple yet effective augmentation approach for the reference signal. During the training phase, \textbf{a random continuous span of the reference signal is masked}. We set the masking ratio as $\theta$. This forces the model to recover the target signal conditioned on a corrupted reference signal, thereby compelling it to \textbf{utilize additional information from the mixture signal}.

Furthermore, we found that distilling from a multi-step model does not significantly improve performance but greatly increases training costs. The performance gains are primarily due to the introduction of the signal quality loss in the single-step model. Therefore, we train the single-step corrector directly from scratch.

\section{Experiments}
\label{exp}
We conducted both in-domain and out-of-domain TSE experiments to evaluate the effectiveness and robustness of SoloSpeech. 
To examine the impact of each component in SoloSpeech, we performed ablation studies on TSE results involving different audio compressors, speaker conditions, and correctors.

\subsection{Datasets}
\label{exp1}

To evaluate the performance of TSE, we trained models on the Libri2Mix 16k Hz dataset \cite{cosentino2020librimix}, which is simulated using WHAM$!$ noise data \cite{DBLP:conf/interspeech/WichernAFZMCMR19} and Librispeech utterances \cite{DBLP:conf/icassp/PanayotovCPK15}. The dataset comprises two training subsets with 212 hours (\textit{train-360}) and 58 hours (\textit{train-100}) of audio, respectively. We utilized the \textit{train-360} subset for training. The \textit{dev} set, containing 11 hours of audio, was used for validation. We then evaluated our models on the \textit{test} set, which includes 11 hours of audio. During inference, the cue speech is a different utterance from the target speech by the same speaker, following the same data preparation as in TD-SpeakerBeam \cite{delcroix2020improving}.
This dataset is also utilized for two-speaker SS experiments.
Although Libri2Mix includes different speakers and background noises in its training and test sets, the recording conditions and speaking styles remain highly similar.
Following \cite{wang2024noise}, we assessed the trained models on out-of-domain data. Utterances from the Wall Street Journal (WSJ) corpus \cite{DBLP:conf/interspeech/IsikRCWH16} served as the speech source, while noise audio from WHAM$!$, MUSAN \cite{DBLP:journals/corr/SnyderCP15}, and DEMAND \cite{DBLP:conf/iwaenc/HadadHVG14} were used as noise sources for simulation. Each of these test sets has a duration of 5 hours. 
Following the approach in \cite{DBLP:conf/interspeech/WichernAFZMCMR19}, noise was introduced by sampling a random SNR value from a uniform distribution ranging from -6 to +3 dB. Additionally, we generated a minimum-length version of the simulated data by removing any leading and trailing noise and truncating it to match the length of the shorter of the two speakers' utterances. During inference, we selected a random speech sample of the target speaker different from the target speech as the cue speech.

We also evaluated the performance on two real-world datasets: CHiME-5 \cite{DBLP:conf/interspeech/BarkerWVT18} and RealSEP \cite{DBLP:journals/corr/abs-2410-01481}.
CHiME-5 consists of real dinner party conversations recorded using distant microphones, featuring up to four overlapping speakers, expressive speech, non-verbal vocalizations (e.g., laughter, sigh), and background sound effects.
RealSEP covers ten diverse real-world acoustic scenes, including music, sound effects, strong reverberation, and moving sound sources.
Both datasets are highly challenging and contain conditions that are significantly different from those in Libri2Mix, demonstrating the robustness of our method in previously unseen scenarios.
For each dataset, we randomly selected 100 mixtures (10–20 seconds) and their corresponding 10-second cue audio samples.

\subsection{Evaluation metrics}
\label{exp2}
To evaluate the performance of SoloSpeech, we conducted TSE experiments using both quality and intelligibility metrics. The perceptual metrics include: (i) Perceptual Evaluation of Speech Quality (PESQ) \cite{DBLP:conf/icassp/RixBHH01}; (ii) Extended Short-Time Objective Intelligibility (ESTOI) \cite{DBLP:journals/taslp/JensenT16}; (iii) Scale-Invariant Signal-to-Noise Ratio (SI-SNR) \cite{DBLP:conf/interspeech/IsikRCWH16}; and (iv) Deep Noise Suppression Mean Opinion Score (DNSMOS) \cite{DBLP:conf/icassp/ReddyGC22}. The first three are intrusive speech perceptual quality metrics, while DNSMOS is a non-intrusive metric that estimates the overall quality of the audio and naturalness.
For the intelligibility metrics, we performed ASR on the speech files using the Whisper \textit{large-v3-turbo} model\footnote{\href{https://huggingface.co/openai/whisper-large-v3-turbo}{https://huggingface.co/openai/whisper-large-v3-turbo}} \cite{DBLP:conf/icml/RadfordKXBMS23} and calculated the Word Error Rate (WER) using the \textit{jiwer} toolkit\footnote{\href{https://github.com/jitsi/jiwer}{https://github.com/jitsi/jiwer}}. In addition, we measured the cosine similarity between the estimated speech and the ground truth using a pretrained WavLM-based speaker verification network\footnote{\href{https://huggingface.co/microsoft/wavlm-base-plus-sv}{https://huggingface.co/microsoft/wavlm-base-plus-sv}} \cite{DBLP:journals/jstsp/ChenWCWLCLKYXWZ22} to assess speaker similarity (SIM).

\begin{table*}[t]
  \caption{Target speech extraction results on the Libri2Mix dataset. $\star$ are the results of models reproduced by us. D/G indicates if the model is discriminative or generative. \textbf{Bold} for the best result and \underline{underline} for the second-best result.}
  \label{tab:result1}
  \centering
  \begin{tabular}{l|c|ccc|c|ccc}
    \toprule
    \multirow{2}{*}{Method} & \multirow{2}{*}{D/G} & \multicolumn{3}{c|}{Perceptual Quality} & Naturalness & \multicolumn{2}{c}{Intelligibility} \\
    & & PESQ $\uparrow$& ESTOI $\uparrow$&SISNR $\uparrow$&DNSMOS $\uparrow$&WER  $\downarrow$&SIM $\uparrow$ \\
    \midrule
    $\star$ TD-SpeakerBeam \cite{delcroix2020improving}& D & $1.66$& $0.70$& $9.21$& $3.14$& $0.21$& $0.93$\\
    DPCCN \cite{han2022dpccn}& D & $1.74$& $0.73$& $9.30$& $3.58$& -& -\\
    $\star$ X-TF-GridNet \cite{DBLP:journals/inffus/HaoLZ24}& D& $1.72$& $0.72$& $9.85$& $3.42$& $0.19$& $0.93$\\
    SUPERB-TSE \cite{10627502} & D& $1.54$& $0.67$& $8.17$& $3.19$& $0.24$& $0.93$\\
    SSL-MHFA \cite{10448315} & D& $1.76$& $\underline{0.74}$& $\underline{10.60}$& $3.22$& $\underline{0.17}$& $\underline{0.94}$\\
    $\star$ USEF-TSE \cite{DBLP:journals/corr/abs-2409-02615}& D& $\underline{1.82}$& $0.72$& $10.17$& $3.48$& $\underline{0.17}$& $\underline{0.94}$\\
    \midrule
    DiscreteTSE \cite{yu2024generation} & G& $1.29$& $0.69$& $-39.95$& -& -& -\\
    DDTSE \cite{zhang2024ddtse}& G&$1.60$ &$0.71$&$7.60$&$\underline{3.74}$&-&-\\
    $\star$ Diff-TSE \cite{kamo23_interspeech}& G& $1.56$& $0.64$& $8.35$& $3.53$& $0.25$& $0.89$ \\
    $\star$ DPM-TSE \cite{hai2024dpm}& G& $1.50$ & $0.60$& $-1.56$& $3.20$& $0.40$& $0.89$ \\
    $\star$ SoloAudio \cite{DBLP:journals/corr/abs-2409-08425}& G & $1.56$ & $0.63$& $2.64$& $3.71$& $0.35$& $0.93$ \\
    \textbf{SoloSpeech (ours)}& G & $\boldsymbol{1.89}$ & $\boldsymbol{0.78}$& $\boldsymbol{11.12}$& $\boldsymbol{3.76}$& $\boldsymbol{0.15}$& $\boldsymbol{0.96}$ \\
    \bottomrule
  \end{tabular}
\end{table*}

For the real-world TSE evaluation, we followed the protocol from \cite{DBLP:journals/corr/abs-2408-16126}, assessing both separation accuracy and perceptual quality through a Mean Opinion Score (MOS) test conducted with 12 human raters recruited via Prolific\footnote{\href{https://www.prolific.com/}{https://www.prolific.com/}}. We randomly selected 100 test samples from the CHiME-5 and RealSEP datasets, ensuring that each sample was annotated by three raters. Each listener rated the separation quality of isolated tracks on a 1–5 scale (1 = Bad, 5 = Excellent), using the original mixture as a reference. 
We report the mean scores along with $95\%$ confidence intervals.


\subsection{Model configurations}
\label{exp3}

\subsubsection{Audio Compressor}
For the audio compressor, the STFT and its inverse (iSTFT) utilize a window size of 640 and a hop size of 320, resulting in latent representations at 50 Hz. The dimensionality $D$ is set to 128. In the encoder, a Conv2D layer is first applied to the complex-valued spectral input, with a kernel size of $3 \times 3$, a zero-padding size of $1 \times 1$, and 128 channels, followed by Group Normalization. Three TF-GridNet blocks are then applied. The embedding dimension for each T-F unit is 128, the kernel size for Unfold and Deconv1D layers is 1, the stride size for Unfold and Deconv1D layers is 1, the number of hidden units in bidirectional LSTMs (BLSTMs) in each direction is 256, the number of output channels in the $1 \times 1$ Conv2D layers to obtain query and key tensors in the self-attention module is 512, and the number of heads in self-attention is 4. After that, we reshape the features and use a Conv1D layer with a channel size of 128. The PReLU function is used for all activations. The decoder is a mirror of the encoder. The model has a total of 49.3 million parameters. 

\textbf{\textit{Loss Functions:}}  
We employed the following loss components:

\begin{itemize}
    \item Reconstruction Loss: Based on a perceptually weighted multi-resolution STFT \cite{steinmetz2020auraloss} with window sizes of $[1280, 640, 320, 160, 80, 40, 20]$ and hop sizes of $[320, 160, 80, 40, 20, 10, 5]$.
    
    \item Adversarial Loss with Feature Matching: Utilizing five convolutional discriminators as described in Encodec \cite{DBLP:journals/tmlr/DefossezCSA23}, this loss employs a fixed mel bin size of 64, window sizes of $[1280, 640, 320, 160, 80]$, and hop sizes of $[320, 160, 80, 40, 20]$. 
    
    \item KL Divergence Loss: Down-weighted by $1 \times 10^{-4}$.
\end{itemize}

The final loss weightings are set to $1.0$ for the multi-resolution STFT loss, $0.1$ for the adversarial loss, and $5.0$ for feature matching. 

During training, we clipped the audio into 100-frame chunks. The VAE was trained using the AdamW optimizer, with a batch size of 16, base learning rate of $1.5 \times 10^{-4}$, while the discriminators used a learning rate of $3 \times 10^{-4}$. The weight decay was set to $1 \times 10^{-3}$, and training was conducted for 10 epochs. It took two days on one NVIDIA A100-80GB GPU.

\subsubsection{Target Extractor}




For the target extractor, we configure the Diffusion Transformer with 12 layers, a model dimension of 768, 12 attention heads, and a hidden-size multiplier of 4, and the Condition Transformer with 12 layers, a model dimension of 384, 6 attention heads, and a hidden-size multiplier of 4. The model has a total of 200.8 million parameters and was trained with a learning rate of $1 \times 10^{-4}$, a weight decay of $1 \times 10^{-4}$, a batch size of 32, and for 200 epochs. The training and inference steps were set to 1000 and 50, respectively, with the variance ranging from $8.5 \times 10^{-4}$ to $1.2 \times 10^{-2}$.
During training, we applied data augmentation by randomly clipping the mixture and cue speech to lengths ranging from 3 seconds up to their original durations. The model was trained on one NVIDIA A100-80GB GPU for six days. 

During inference, we found that classifier-free guidance (CFG) \cite{DBLP:journals/corr/abs-2207-12598} did not improve performance in TSE. Large guidance scales even tended to produce overly smoothed outputs, with reduced speech detail and degraded intelligibility. As a result, we chose to exclude CFG in our pipeline.
Instead, following \cite{saharia2022photorealistic}, we generate 8 candidates and select the one with the highest speaker similarity score\footnote{\href{https://huggingface.co/microsoft/wavlm-base-plus-sv}{https://huggingface.co/microsoft/wavlm-base-plus-sv}}.

\subsubsection{Corrector}

\begin{table*}[t]
  \caption{Target speech extraction results on the out-of-domain data. We trained the models on the Libri2Mix dataset and then tested them on these datasets.}
  \label{tab:resultreal}
  \centering
  \begin{tabular}{l|l|cccc}
    \toprule
    DataSet & Method & SISNR$\uparrow$ & WER$\downarrow$ & DNSMOS $\uparrow$&MOS $\uparrow$ \\
    \midrule
    \multirow{3}{*}{WHAM!} & USEF-TSE \cite{DBLP:journals/corr/abs-2409-02615}&$9.15$&$0.23$&$3.19$&-\\
    & SoloAudio \cite{DBLP:journals/corr/abs-2409-08425}& $2.52$ & $0.36$& $3.45$&- \\
    & \textbf{SoloSpeech (ours)}& $\boldsymbol{10.68}$& $\boldsymbol{0.18}$& $\boldsymbol{3.72}$&- \\
    \midrule
    \multirow{3}{*}{MUSAN} & USEF-TSE \cite{DBLP:journals/corr/abs-2409-02615}& $8.96$& $0.25$& $0.20$&-\\
    & SoloAudio \cite{DBLP:journals/corr/abs-2409-08425}& $2.05$& $0.38$& $3.38$&- \\
    & \textbf{SoloSpeech (ours)}& $\boldsymbol{10.40}$& $\boldsymbol{0.20}$& $\boldsymbol{3.70}$&- \\
    \midrule
    \multirow{3}{*}{DEMAND} & USEF-TSE \cite{DBLP:journals/corr/abs-2409-02615}& $9.70$& $0.19$& $3.25$&-\\
    & SoloAudio \cite{DBLP:journals/corr/abs-2409-08425}& $3.30$& $0.34$& $3.50$&- \\
    & \textbf{SoloSpeech (ours)}& $\boldsymbol{11.41}$& $\boldsymbol{0.17}$& $\boldsymbol{3.72}$&- \\
    \midrule
    \multirow{3}{*}{CHiME-5 (real-world)} & USEF-TSE \cite{DBLP:journals/corr/abs-2409-02615}&-&- & $2.75$ & $2.10\pm0.15$ \\
    & SoloAudio \cite{DBLP:journals/corr/abs-2409-08425} &-&-& $2.90$ & $2.26\pm0.14$ \\
    & \textbf{SoloSpeech (ours)} &-&-& $\boldsymbol{3.38}$ & $\boldsymbol{2.93\pm0.18}$ \\
    \midrule
    \multirow{3}{*}{RealSEP (real-world)} & USEF-TSE \cite{DBLP:journals/corr/abs-2409-02615}&-&- & $2.69$ & $2.04\pm0.15$ \\
    & SoloAudio \cite{DBLP:journals/corr/abs-2409-08425} &-&-& $2.52$ & $1.87\pm0.20$ \\
    & \textbf{SoloSpeech (ours)} &-&-& $\boldsymbol{3.15}$ & $\boldsymbol{2.70\pm0.18}$ \\
    \bottomrule
  \end{tabular}
\end{table*}

For the corrector, we employed the same network as Fast-GeCo \cite{wang2024noise}, which is based on a multi-resolution U-Net structure operating on complex-valued STFT, and set the reverse starting point $\tau = 0.5$. The STFT utilizes a window size of 510 and a hop size of 128. During training, we clipped the audio into 256-frame chunks.
We used a learning rate of $1 \times 10^{-4}$, a batch size of 16 and 50 epochs, which took three days on one NVIDIA A100-80GB GPU.

\subsection{Baselines}
\begin{itemize}
    \item \textbf{TD-SpeakerBeam} \cite{delcroix2020improving}: A time-domain discriminative model that investigates strategies for improving speaker discrimination capability. We utilized their open-source code\footnote{\href{https://github.com/BUTSpeechFIT/speakerbeam}{https://github.com/BUTSpeechFIT/speakerbeam}}.
    \item \textbf{DPCCN} \cite{han2022dpccn}: A densely-connected pyramid complex convolutional network serving as a time-frequency domain discriminative model. We used the results reported in \cite{zhang2024ddtse}.
    \item \textbf{SUPERB-TSE} \cite{10627502}: A discriminative model based on large-scale pre-trained SSL models. We reached out to the authors to obtain the test results.
    \item \textbf{SSL-MHFA} \cite{10448315}: A discriminative model based on large-scale pre-trained SSL models with Adaptive Input Enhancer (AIE) and a speaker encoder.
    The test results were requested from the authors.
    \item \textbf{X-TF-GridNet} \cite{DBLP:journals/inffus/HaoLZ24}: a time-frequency domain discriminative model using TF-GridNet as its backbone. It extracts speaker embeddings using a U$^2$-Net style network. We used their open-source code\footnote{\href{https://github.com/HaoFengyuan/X-TF-GridNet}{https://github.com/HaoFengyuan/X-TF-GridNet}}.
    \item \textbf{USEF-TSE} \cite{DBLP:journals/corr/abs-2409-02615}: A speaker embedding-free discriminative model. We implemented the time-frequency model, \textit{i.e.}, USEF-TFGridNet, as described in the work. We used their open-source code\footnote{\href{https://github.com/ZBang/USEF-TSE}{https://github.com/ZBang/USEF-TSE}}.
    \item \textbf{DiscreteTSE} \cite{yu2024generation}: A discrete token-based TSE approach that combines speech discretization and vocoder techniques. We used the results reported in the paper.
    \item \textbf{DDTSE} \cite{zhang2024ddtse}: A generative model that applies the same forward process as diffusion models and uses a reconstruction loss similar to discriminative methods. We used the results reported in the paper.
    \item \textbf{Diff-TSE} \cite{kamo23_interspeech}: A time-frequency domain generative model based on a conditional diffusion model. Since no open-source code is available, we reproduced the results based on the paper.
    \item \textbf{DPM-TSE} \cite{hai2024dpm}: A diffusion probabilistic model operating on the mel-spectrogram. We utilized a pretrained WavLM-based speaker verification network as the speaker embedding network. We used their open-source code\footnote{\href{https://github.com/haidog-yaqub/DPMTSE}{https://github.com/haidog-yaqub/DPMTSE}}. 
    \item \textbf{SoloAudio} \cite{DBLP:journals/corr/abs-2409-08425}: A latent diffusion model with a time-domain audio VAE as the latent features. We utilized a pretrained WavLM-based speaker verification network as the speaker embedding network. We utilized their open-source code\footnote{\href{https://github.com/WangHelin1997/SoloAudio}{https://github.com/WangHelin1997/SoloAudio}}.
\end{itemize}

\subsection{Comparisons with state-of-the-art methods}
\label{exp4}
To evaluate the efficiency of the proposed approach, we compared SoloSpeech with existing methods. 
We reproduced the results using the authors' official settings. If the authors employed the same dataset as us, we report their metric values as presented in their original papers.

\subsubsection{In-domain Results} We trained and tested the models on the Libri2Mix dataset. As shown in Table~\ref{tab:result1}, SoloSpeech consistently outperformed existing methods across all metrics, including perceptual quality, naturalness, and intelligibility, achieving a 0.95 dB gain in SI-SNR.
Compared to discriminative methods, SoloSpeech achieves significant improvements in DNSMOS, demonstrating its powerful ability to produce natural and high-quality audio. Furthermore, when compared with other generative approaches, SoloSpeech significantly outperforms them in perceptual and intelligibility metrics, highlighting its advantages.
Higher values of DNSMOS and lower WER also indicate that our method produces fewer artifacts for both listeners and ASR systems.

\subsubsection{Out-of-domain Results} We trained the models on the Libri2Mix dataset and tested them on three out-of-domain datasets \textbf{without any fine-tuning}. As shown in Table~\ref{tab:resultreal}, SoloSpeech consistently outperforms existing methods in generalizing to unseen conditions. Specifically, SoloSpeech achieves SI-SNR gains of 16.7\%, 16.1\%, and 17.6\% over the state-of-the-art discriminative method USEF-TSE on the WHAM!, MUSAN, and DEMAND datasets, respectively.

\begin{table*}[t]
  \caption{Impact of the modules. Unless specific statement, corrector is not applied. $\ast$ used fixed speaker embedding extracted from a pretrained WavLM-based speaker verification model. $\bullet$ finetuned the speaker embedding. $\diamond$ used features from a WavLM model.}
  \label{tab:ablation}
  \centering
  \begin{tabular}{l|l|ccc}
    \toprule
    Module & Method &SISNR $\uparrow$&DNSMOS $\uparrow$&WER $\downarrow$ \\
    \midrule
    \multirow{2}{*}{Compressor} & Stable Audio VAE \cite{DBLP:journals/corr/abs-2407-14358} &$9.37$ & $3.69$ & $0.20$ \\
    &\textbf{T-F Audio VAE (ours)} & $\boldsymbol{10.10}$& $\boldsymbol{3.76}$& $\boldsymbol{0.18}$ \\
    \midrule
    \multirow{5}{*}{Speaker Condition} &$\ast$ Speaker embedding \cite{DBLP:journals/jstsp/ChenWCWLCLKYXWZ22} & $9.25$ & $3.72$ & $0.20$  \\
    &$\bullet$ Speaker embedding
 \cite{DBLP:journals/jstsp/ChenWCWLCLKYXWZ22} & $8.77$ & $3.65$ & $0.22$  \\
    &$\diamond$ SSL-MHFA \cite{10448315} &$8.85$ & $3.70$& $0.22$\\
    &Speaker network \cite{delcroix2020improving} &$8.32$ & $3.69$& $0.26$\\
    &\textbf{Latent-space fusion (ours)} &$\boldsymbol{10.10}$ & $\boldsymbol{3.76}$& $\boldsymbol{0.18}$ \\
    \midrule
    \multirow{5}{*}{Corrector} & USEF-TSE \cite{DBLP:journals/corr/abs-2409-02615} &$10.17$ & $3.48$& $0.17$   \\
    &USEF-TSE w/ SoloSpeech corrector &$11.06$ & $3.55$& $0.16$   \\
    &SoloAudio \cite{DBLP:journals/corr/abs-2409-08425} &$2.64$ & $3.71$& $0.35$   \\
    &SoloAudio w/ SoloSpeech corrector &$6.58$ & $3.72$& $0.22$   \\
    &SoloSpeech w/o corrector &$10.10$ & $\boldsymbol{3.76}$& $0.18$\\
    &SoloSpeech w/ Fast-GeCo corrector \cite{wang2024noise} &$10.55$ & $\boldsymbol{3.76}$& $0.17$\\
    &\textbf{SoloSpeech (ours)} &$\boldsymbol{11.12}$ & $\boldsymbol{3.76}$& $\boldsymbol{0.15}$\\
    \bottomrule
  \end{tabular}
\end{table*}

\subsubsection{Real-world Results} We trained models on the Libri2Mix and tested them on two real-world datasets \textbf{without any fine-tuning}. As shown in Table~\ref{tab:resultreal}, SoloSpeech significantly outperforms baseline methods in
generalizing to real-world challenging scenarios. Remarkably, SoloSpeech is able to generalize to more challenging unseen conditions. For example, handling more overlapping speakers (up to 4 in CHiME-5 vs. 2 in Libri2Mix) and much longer audio durations (up to 20 seconds).
Moreover, it demonstrates strong robustness to factors such as expressive speech, non-verbal vocalizations, reverberation, and background sound effects.

\begin{table}[t]
  \caption{Reconstruction results of different audio compressors on the Libri2Mix dataset.}
  \label{tab:result4}
  \centering
  \begin{tabular}{l|l|ccc}
    \toprule
    Data& Method &PESQ&ESTOI&SISNR\\
    \midrule
    \multirow{2}{*}{Clean Single} & 
    Stable Audio VAE 
    \cite{DBLP:journals/corr/abs-2407-14358}& $4.151$& $0.984$& $17.094$ \\
    & T-F Audio VAE (ours)  & $\boldsymbol{4.506}$& $\boldsymbol{0.995}$& $\boldsymbol{18.379}$ \\
    \midrule
    \multirow{2}{*}{Noisy mixture} & 
    Stable Audio VAE \cite{DBLP:journals/corr/abs-2407-14358}& $4.327$& $0.984$& $14.575$ \\
    & T-F Audio VAE (ours)  & $\boldsymbol{4.529}$& $\boldsymbol{0.995}$& $\boldsymbol{15.571}$ \\
    \bottomrule
  \end{tabular}
\end{table}

\subsection{Ablation studies}
\label{exp5}
To gain a deeper understanding of SoloSpeech, we compared each key component using the Libri2Mix dataset under a completely fair setup, employing identical architectures and hyperparameters in the subsequent experiments.

\subsubsection{Comparison of audio compressors}
In the audio compressors reconstruction experiments, we utilized a set of audio quality metrics: PESQ, ESTOI, and SI-SNR. As shown in Table~\ref{tab:result4}, the T-F Audio VAE consistently outperforms the Stable Audio VAE across all metrics, demonstrating superior audio reconstruction quality.

\begin{table}[t]
  \caption{Impact of different settings for the audio compressor on the noisy mixture.}
  \label{tab:vae}
  \centering
  \begin{tabular}{l|c|ccc}
    \toprule
Frame Rate & Latent Dim &PESQ &ESTOI &SISNR\\
    \midrule
    \multirow{2}{*}{25 Hz} & 64 & $3.979$& $0.978$& $14.105$  \\
    &128 & $4.204$& $0.988$& $14.996$ \\
    \midrule
    \multirow{2}{*}{50 Hz} & 64& $4.268$& $0.989$& $15.080$   \\
    &128& $4.529$& $0.995$& $15.571$ \\
    \midrule
    \multirow{2}{*}{100 Hz} & 64 & $4.510$& $0.995$& $15.499$  \\
    &128& $4.530$& $0.995$& $15.688$ \\
    \bottomrule
  \end{tabular}
\end{table}

We conducted experiments to evaluate the reconstruction quality under different audio compressor settings, including variations in frame rate and latent dimension. Table~\ref{tab:vae} summarizes the impact of these configurations. Lower frame rates and smaller latent dimensions generally make it easier to train latent diffusion models. Ultimately, we select a frame rate of 50 Hz and a latent dimension of 128 as a trade-off between training efficiency and reconstruction quality.

In addition, we compared the proposed T-F Audio VAE with the state-of-the-art Stable Audio VAE. We employed the VAEs as audio compressors within SoloSpeech and assessed their impact on the TSE task. As presented in Table~\ref{tab:ablation}, the newly proposed T-F Audio VAE significantly enhances TSE performance compared to the Stable Audio VAE.

\subsubsection{Comparison of speaker conditions} We compared different speaker conditions and reported the results in Table~\ref{tab:ablation}. Specifically, We evaluated a speaker embedding\footnote{\href{https://github.com/TaoRuijie/ECAPA-TDNN}{https://github.com/TaoRuijie/ECAPA-TDNN}}-based approach with AdaLN \cite{DBLP:journals/corr/abs-2409-08425}, considering both fixed and fine-tuned configurations; an SSL-based speaker network\footnote{\href{https://github.com/BUTSpeechFIT/wespeaker_ssl_public}{https://github.com/BUTSpeechFIT/wespeaker\_ssl\_public}} \cite{10448315}; a time-domain speaker network\footnote{\href{https://github.com/BUTSpeechFIT/speakerbeam}{https://github.com/BUTSpeechFIT/speakerbeam}}
\cite{delcroix2020improving}; and our proposed VAE latent-domain speaker network. Our method does not require additional data to train speaker embeddings, yet achieves the best results among all methods. This superiority is attributed to our approach's ability to preserve the sequential information of the conditional features and prevent potential mismatching between VAE latents and other types of features representing the target speech. 

\begin{figure}[t]
\begin{center}
\centerline{\includegraphics[width=0.9\columnwidth]{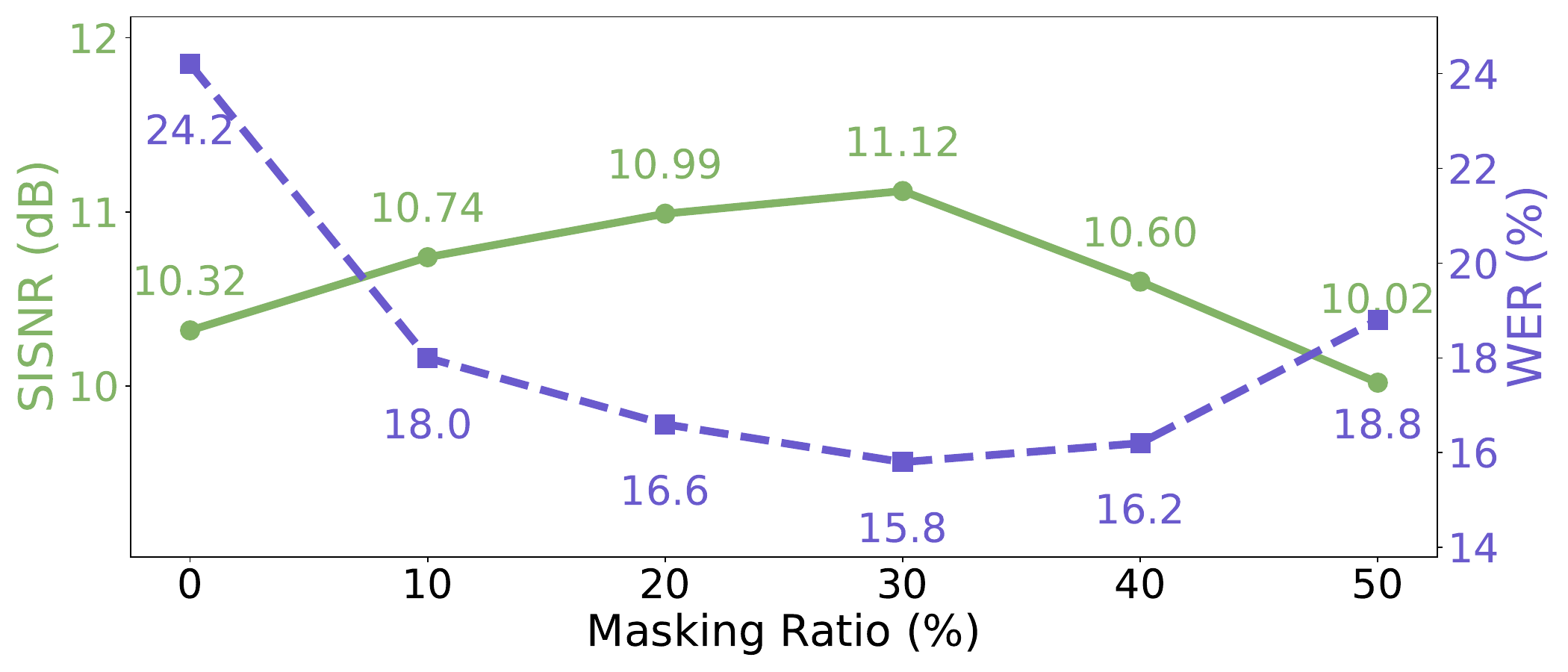}}
\caption{Results of different masking ratios ($\theta$) for the corrector.}
\label{pic:mask}
\end{center}
\end{figure}

\subsubsection{Impact of the corrector} 
We investigated the impact of the corrector and present the results in Table~\ref{tab:ablation}. Our proposed corrector significantly outperforms the previous work, Fast-GeCo, thereby substantially enhancing perceptual quality and intelligibility.
The corrector contributes significantly to SI-SNR and WER gains, as we found that it can significantly improve signal quality and reduce artifacts in high frequencies, which may influence the ASR system. 
Furthermore, we applied the proposed corrector to two prior baselines, USEF-TSE (a discriminative method) and SoloAudio (a generative method), and reported the results in Table~\ref{tab:ablation}. The corrector consistently improved performance across all evaluation metrics.
Notably, SoloSpeech still outperforms USEF-TSE even with the corrector applied, especially in DNSMOS scores, highlighting the advantage of generative models in producing more natural-sounding speech.
Compared to SoloAudio with the same corrector, SoloSpeech achieves significantly better perceptual quality and intelligibility, demonstrating the effectiveness of our target extractor backbone.

\subsubsection{Comparison of Masking ratios}
We explored the influence of different masking ratios ranging from 0\% to 50\% during the training of the corrector. We measured SI-SNR for perceptual quality (higher values indicate better quality) and WER for intelligibility (lower values indicate better performance). As shown in Figure~\ref{pic:mask}, increasing the masking ratio initially improves overall performance but eventually degrades it, resulting in an optimal ratio of 30\%. We argue that the masking ratio determines the learning dynamics of the corrector: a small ratio causes the model to focus more on the reference audio, while a high ratio leads the model to consider more of the mixture audio. Striking a balance between these factors yields the best perceptual quality and intelligibility for the TSE task.

\subsubsection{Comparison of audio durations}
SoloSpeech was trained on audio segments of varying lengths, ranging from 3 to 10 seconds. The model also supports longer input thanks to the Rotary Position Embeddings (RoPE) in the Transformer backbone. In our experiments on Libri2Mix, CHiME-5, and RealSEP, the input segments range from 3 to 20 seconds, and we observe no significant degradation in performance as the input length increases, which is shown in Table~\ref{tab:result10}. This indicates that SoloSpeech is robust to variations in audio duration.

\begin{table}[t]
  \caption{Comparison of different audio durations.}
  \label{tab:result10}
  \centering
  \begin{tabular}{l|ccc}
    \toprule
Audio Length (s) &SISNR $\uparrow$&DNSMOS $\uparrow$&WER $\downarrow$\\
    \midrule
    2.5-3.5 & $11.08$&$3.79$&$0.18$ \\
    5.5-6.5 & $11.16$ &$3.73$&$0.16$\\
    $>$ 9 & $11.20$ &$3.70$&$0.15$\\
    all & $11.12$ &$3.76$&$0.16$\\
    \bottomrule
  \end{tabular}
\end{table}

\section{Visualization analysis}
\label{ap10}
\begin{figure*}[t]
  \centering
  \subfloat[]{\includegraphics[width=0.3\linewidth]{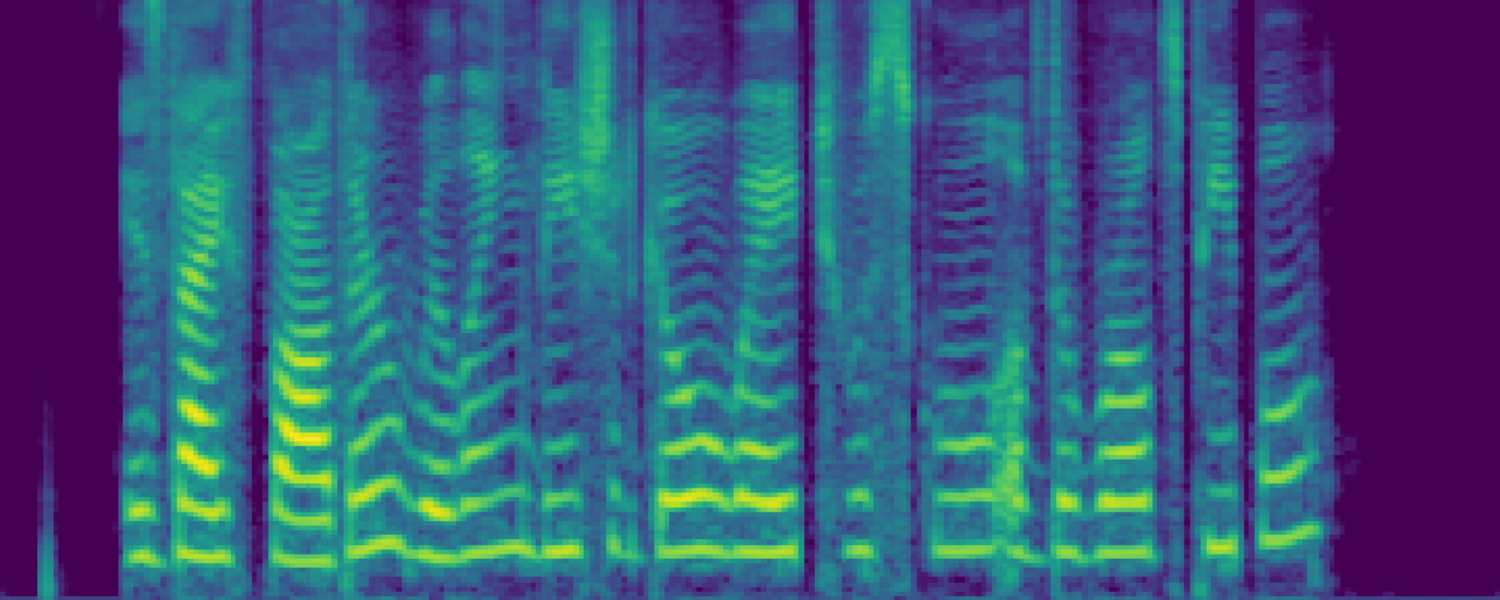}\label{subfig:1}}
  \hfill
  \subfloat[]{\includegraphics[width=0.3\linewidth]{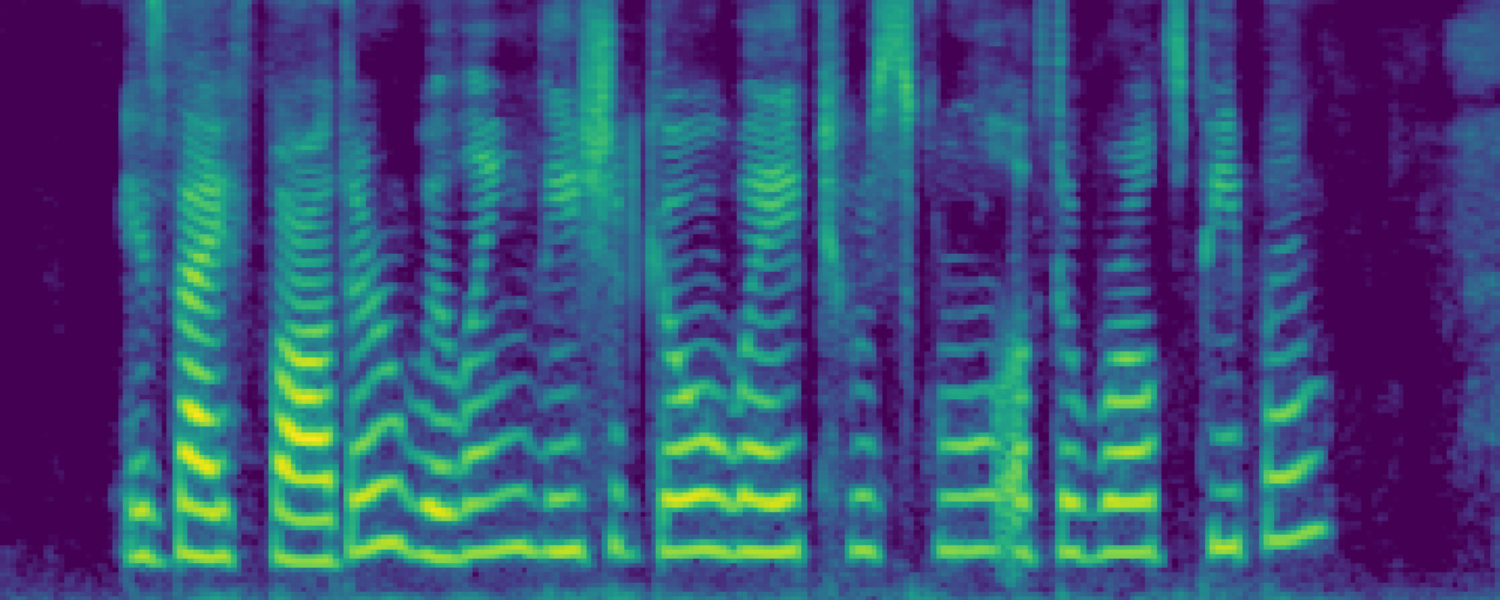}\label{subfig:2}}
  \hfill
  \subfloat[]{\includegraphics[width=0.3\linewidth]{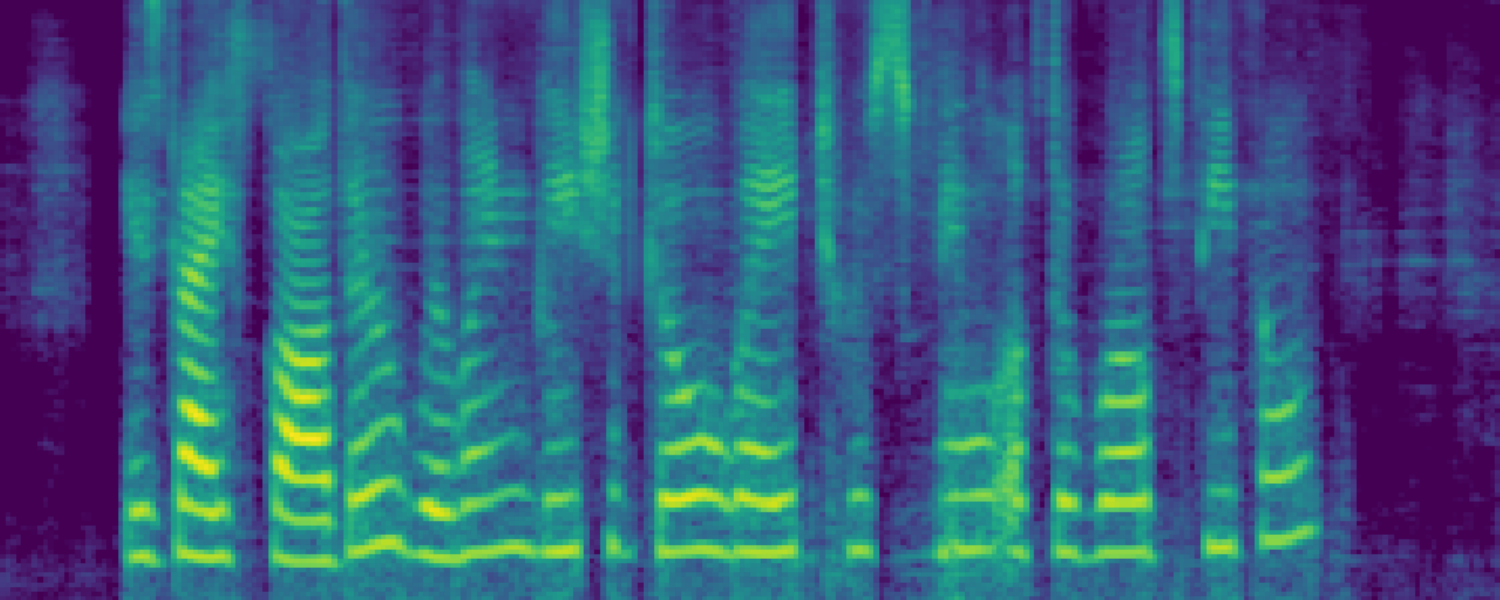}\label{subfig:3}}
  
  
  \subfloat[]{\includegraphics[width=0.3\linewidth]{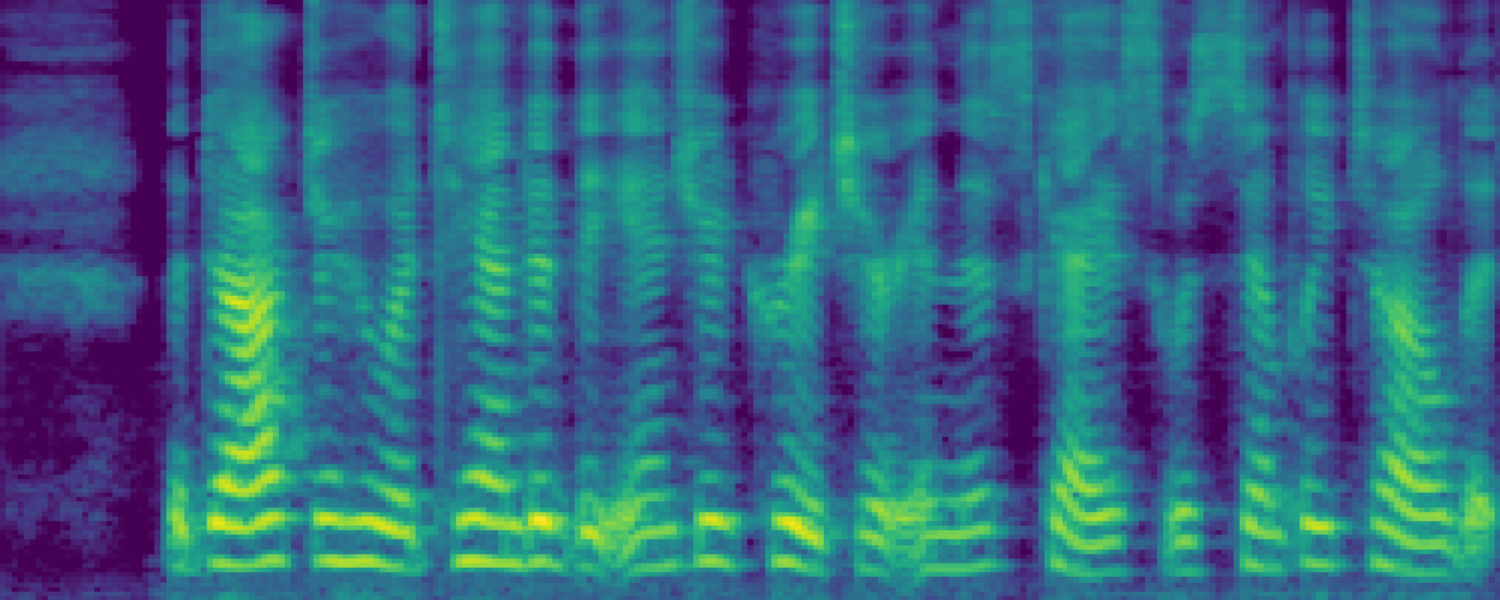}\label{subfig:4}}
  \hfill
  \subfloat[]{\includegraphics[width=0.3\linewidth]{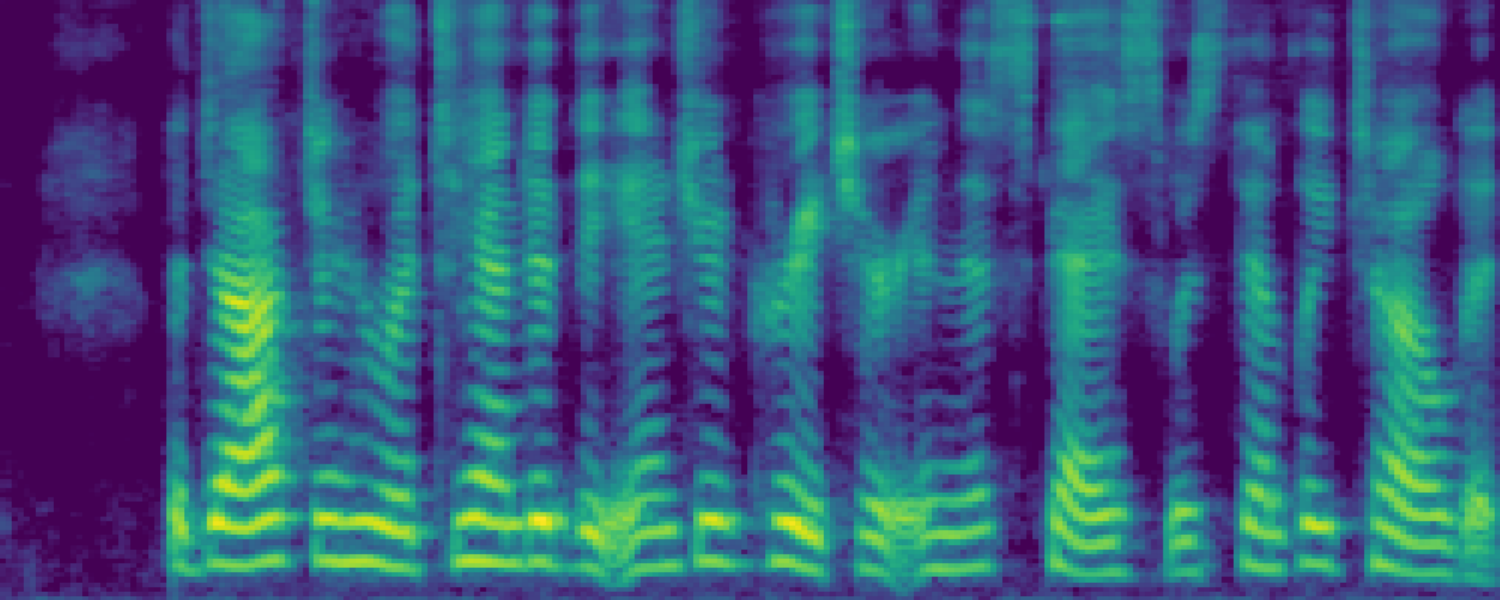}\label{subfig:5}}
  \hfill
  \subfloat[]{\includegraphics[width=0.3\linewidth]{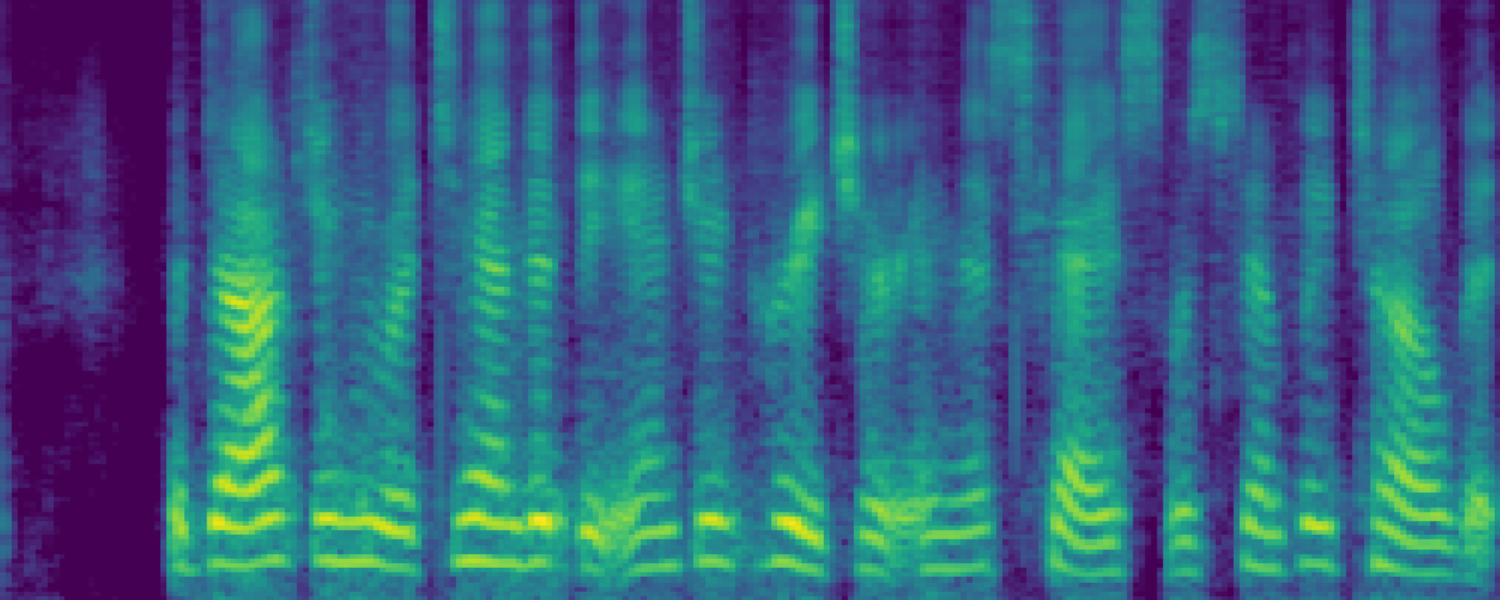}\label{subfig:6}}
  
  
  \subfloat[]{\includegraphics[width=0.3\linewidth]{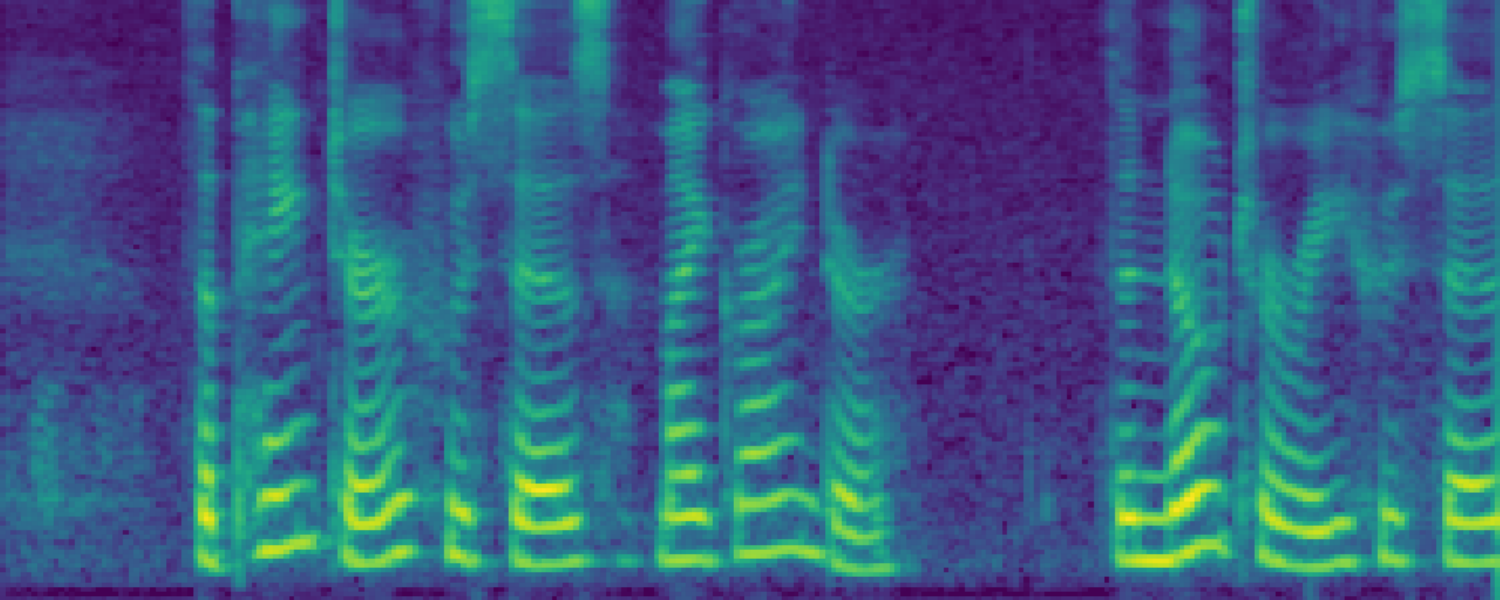}\label{subfig:7}}
  \hfill
  \subfloat[]{\includegraphics[width=0.3\linewidth]{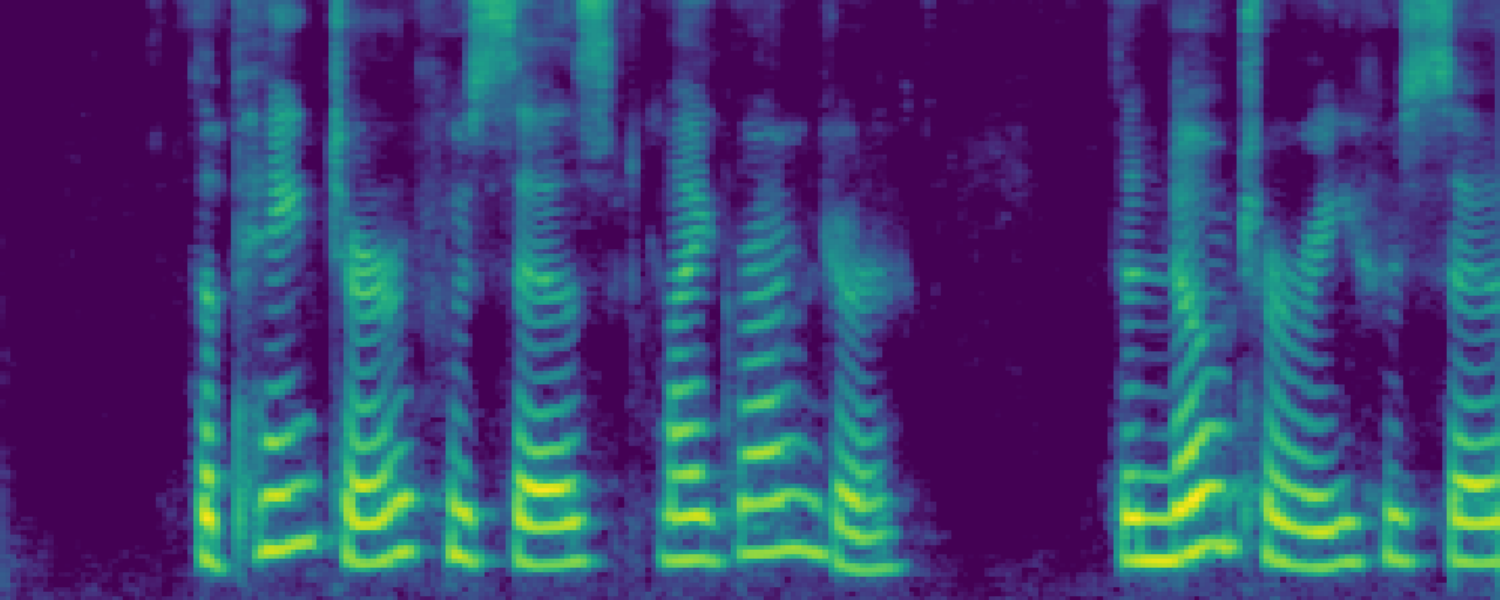}\label{subfig:8}}
  \hfill
  \subfloat[]{\includegraphics[width=0.3\linewidth]{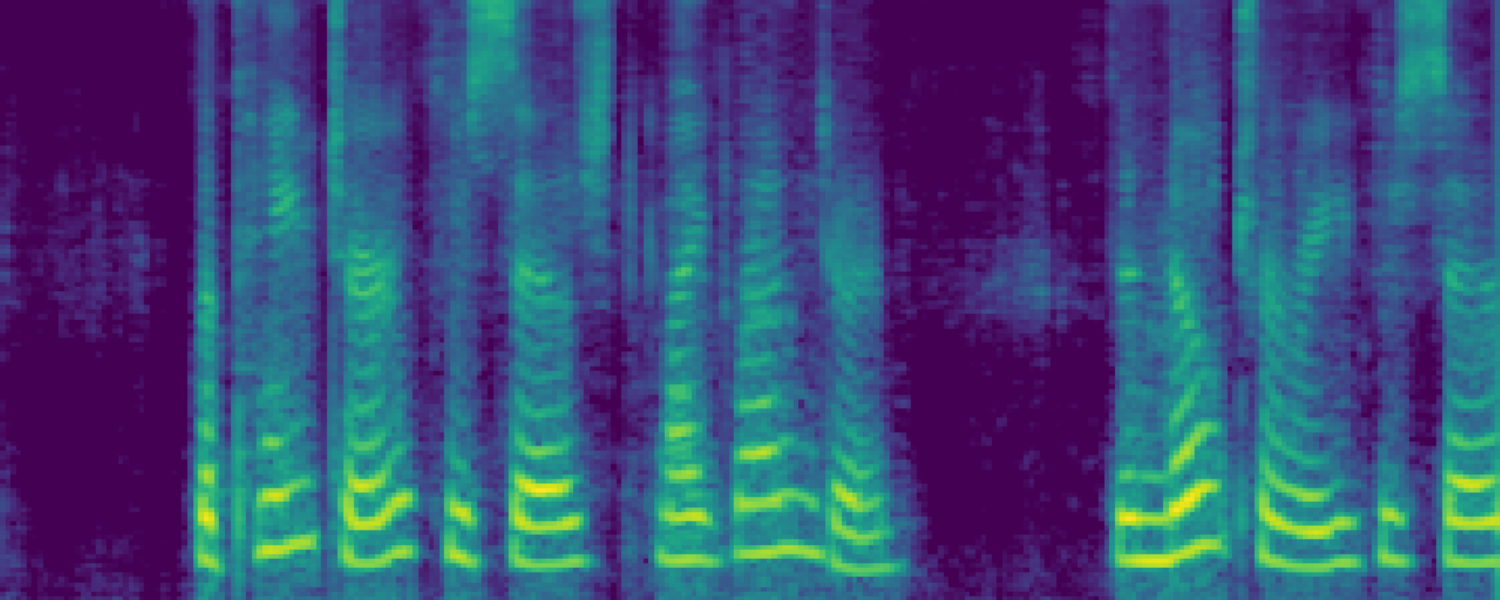}\label{subfig:9}}
  
  
  \subfloat[]{\includegraphics[width=0.3\linewidth]{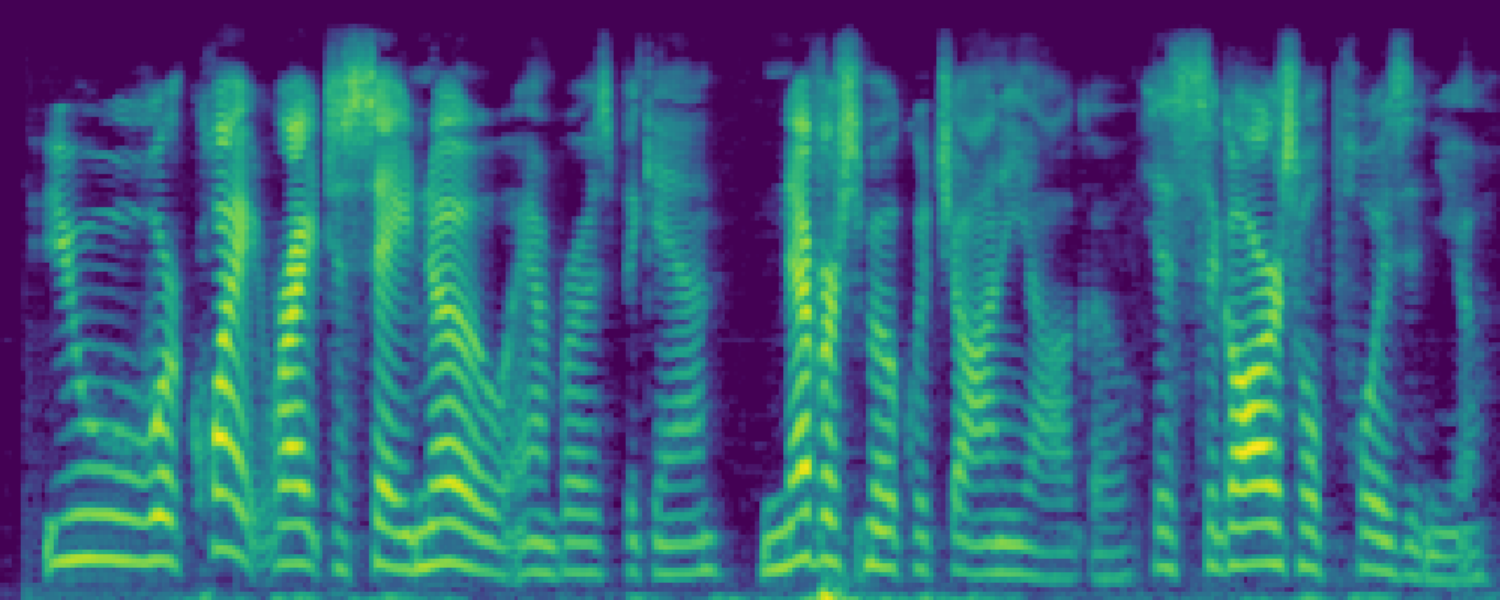}\label{subfig:10}}
  \hfill
  \subfloat[]{\includegraphics[width=0.3\linewidth]{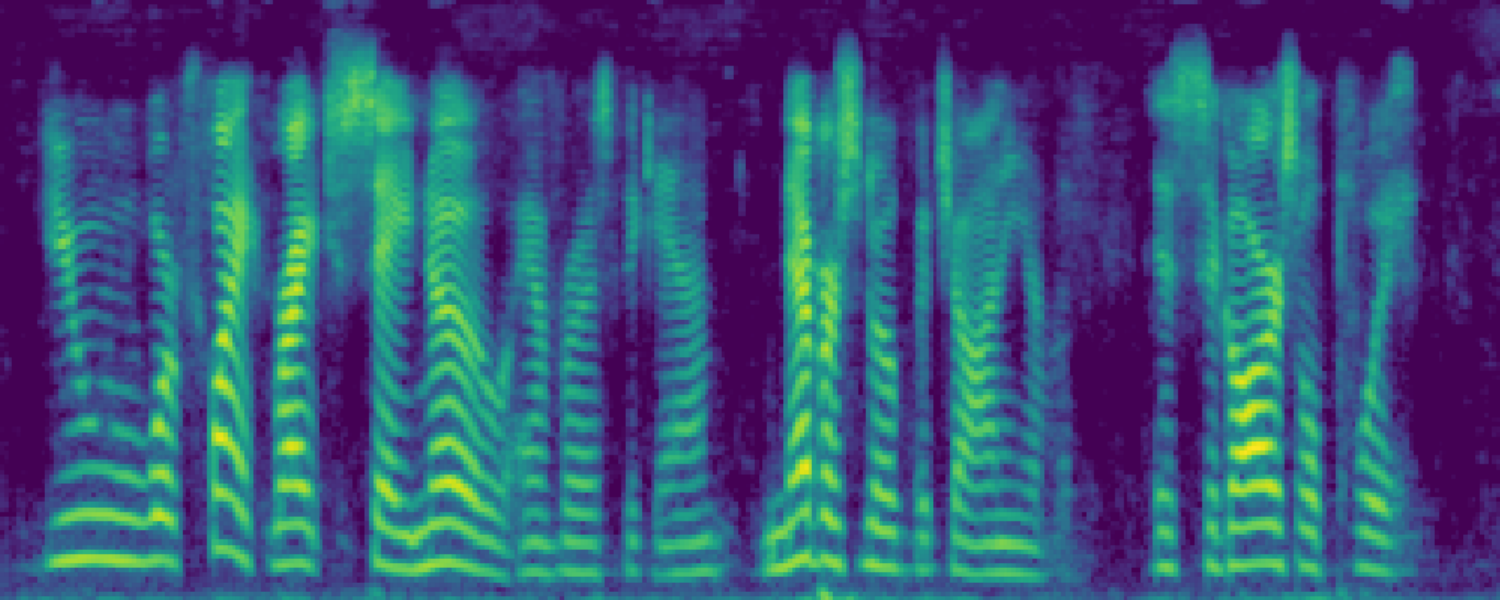}\label{subfig:11}}
  \hfill
  \subfloat[]{\includegraphics[width=0.3\linewidth]{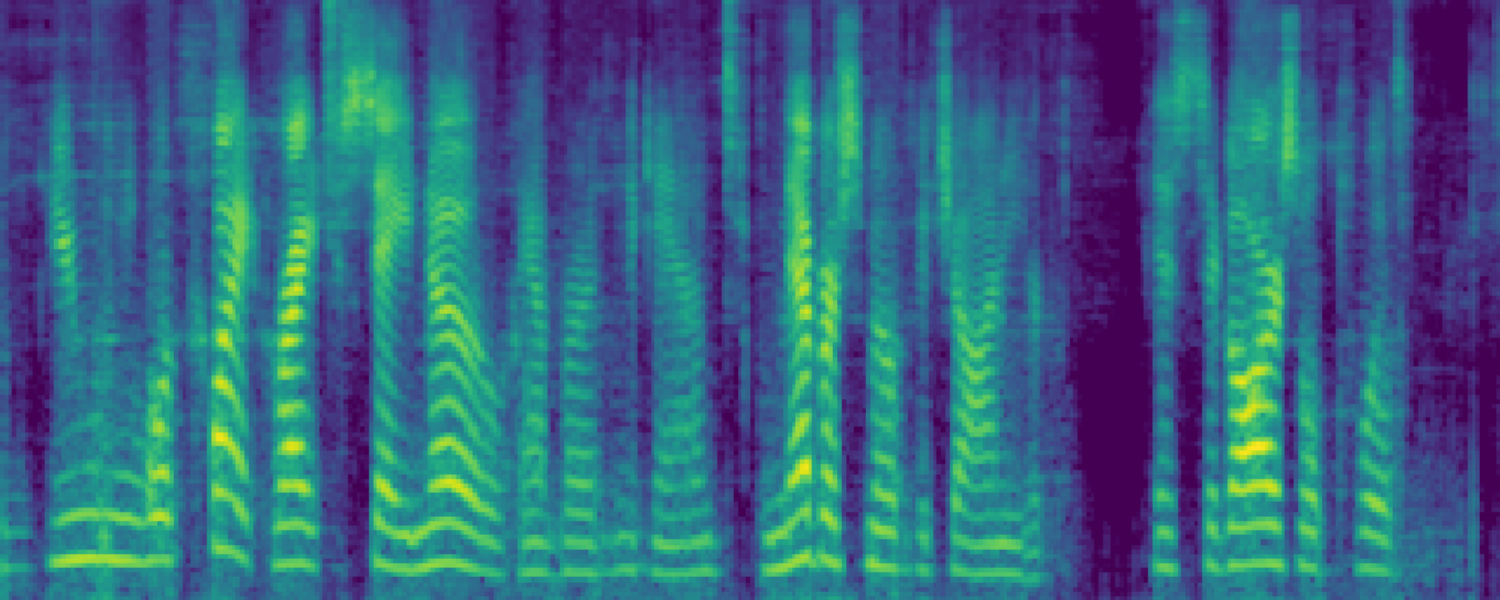}\label{subfig:12}}
  
  
  \subfloat[]{\includegraphics[width=0.3\linewidth]{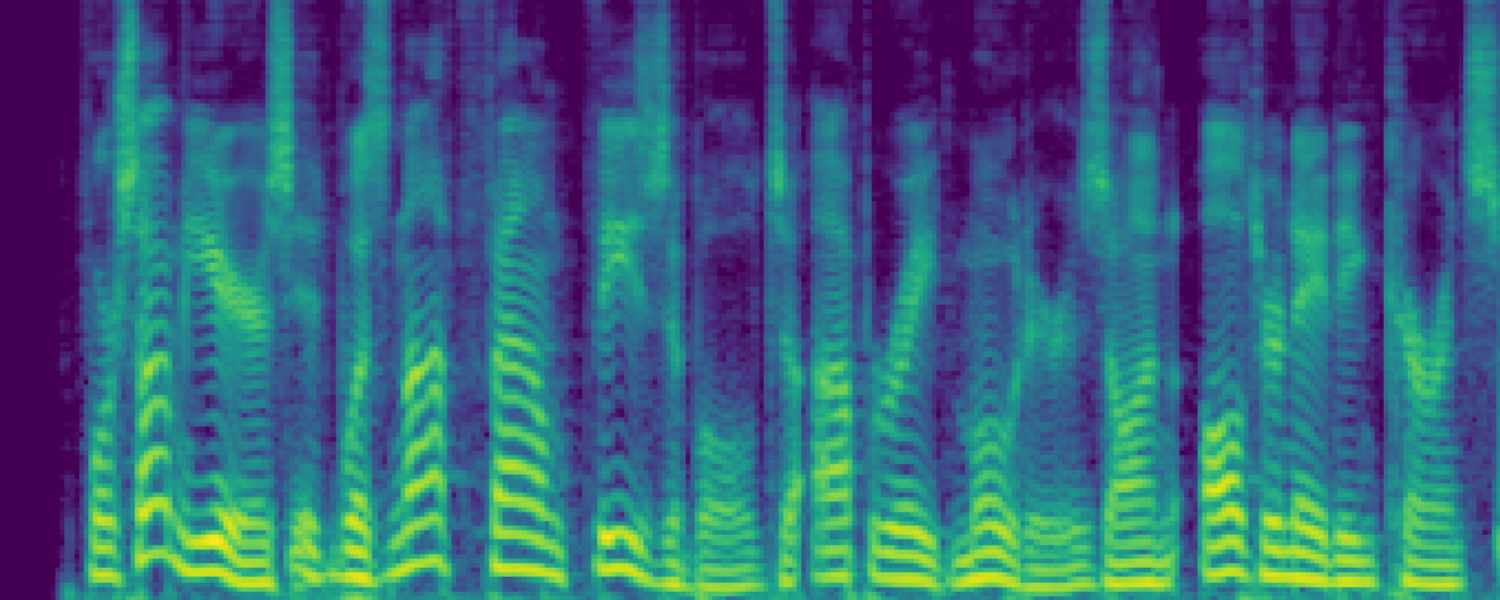}\label{subfig:13}}
  \hfill
  \subfloat[]{\includegraphics[width=0.3\linewidth]{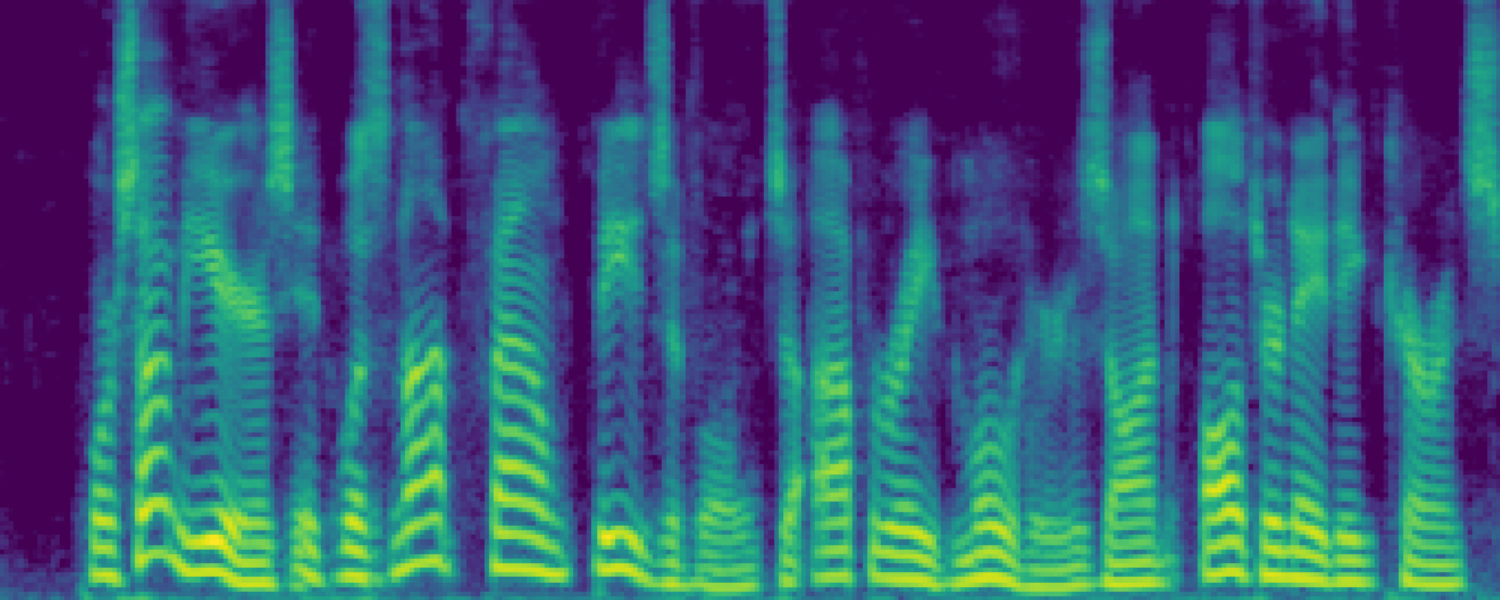}\label{subfig:14}}
  \hfill
  \subfloat[]{\includegraphics[width=0.3\linewidth]{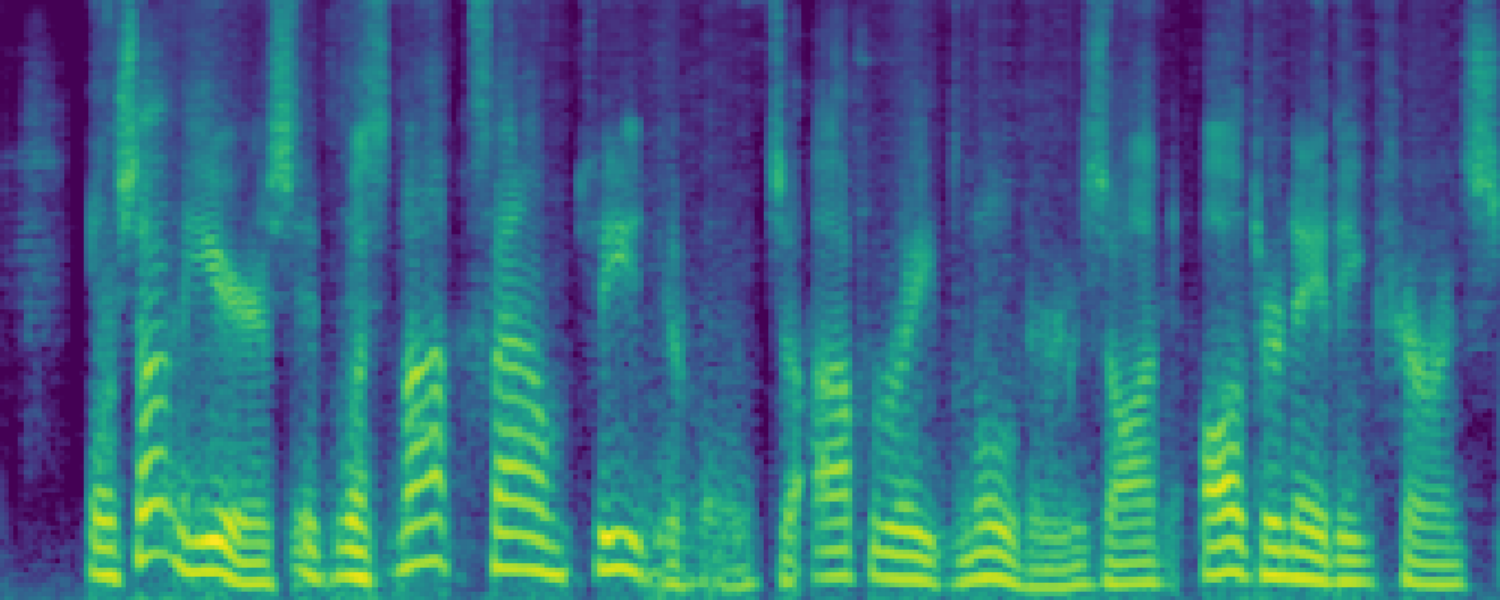}\label{subfig:15}}

  \caption{Comparison of the spectrograms of the ground truth, audio extracted by SoloSpeech, and by USEF-TSE.
  Rows: Sample I--V. 
  Columns: (a,d,g,j,m) Ground truth, (b,e,h,k,n) audio extracted by SoloSpeech, (c,f,i,l,o) audio extracted by USEF-TSE.}
  \label{fig:grid}
\end{figure*}

To intuitively showcase the extraction performance of SoloSpeech, we provide several visualization examples illustrated in Figure \ref{fig:grid}. The spectrograms below compare the inference results of SoloSpeech and USEF-TSE applied to the same audio input, alongside the ground truth. All samples demonstrate that SoloSpeech achieves finer reconstruction, particularly at high frequencies, compared to USEF-TSE. In addition, SoloSpeech exhibits superior noise reduction and effectively prevents spectrum leakage, as evidenced in Samples IV and V.
Details of these samples are as follows:

\begin{itemize} 
    \item \textbf{Sample I}: SoloSpeech achieves better noise reduction and reconstructs high frequencies more accurately. In contrast, USEF-TSE introduces artifacts at the beginning and end of the speech. \item \textbf{Sample II}: SoloSpeech provides superior reconstruction of high frequencies and better recovery of the speech onset. 
    \item \textbf{Sample III}: SoloSpeech produces fewer artifacts, resulting in cleaner speech. 
    \item \textbf{Sample IV}: SoloSpeech reconstructs high frequencies much better and introduces fewer artifacts in all the frequencies. \item \textbf{Sample V}: USEF-TSE introduces some artifacts at the beginning of the speech and results in more noise across all frequency bands.
\end{itemize}



\section{Conclusion and Discussions}
\label{conclusions}
We proposed a cascaded generative pipeline that offers a strong foundation for future research in generative TSE. 
Each component is validated for its effectiveness and designed with modularity and scalability in mind.

Compared to previous state-of-the-art methods in both TSE and SS tasks, SoloSpeech achieves significantly higher perceptual quality, naturalness, and intelligibility. 
Furthermore, when evaluated on out-of-domain data, SoloSpeech demonstrates excellent generalization and robustness.

In real-world evaluations, we observe that strong reverberation and moving sound sources, such as those in the RealSEP dataset, remain challenging, and we leave these issues to future work. We also plan to explore more efficient backbone architectures for each component.

\bibliographystyle{IEEEtran}

\bibliography{ref}











\newpage

 





\end{document}